\documentclass[titlepage,12pt,subeqn]{utarticle}
\usepackage{amsmath,amssymb,amscd}
\usepackage{color}
\usepackage{latexsym}
\usepackage{ifthen}
\usepackage[xdvi]{graphics}
\usepackage[]{hyperref} %uncomment for HyperTeX links
\usepackage{bbm}

\numberwithin{equation}{section}

\newcommand{\Tr}{{\rm Tr}}

\newcommand{\ic}{{\rm i}}

\newcommand{\be}{\begin{equation}}
\newcommand{\ee}{\end{equation}}
\newcommand{\bea}{\begin{eqnarray}}
\newcommand{\eea}{\end{eqnarray}}
\newcommand{\al}{\alpha}
\renewcommand{\d}{\delta}
\newcommand{\e}{\epsilon}
\newcommand{\G}{\Gamma}
\newcommand{\g}{\gamma}
\renewcommand{\k}{\kappa}

\newcommand{\m}{\mu}
\newcommand{\n}{\nu}
\newcommand{\Om}{\Omega}
\newcommand{\om}{\omega}
\newcommand{\s}{\sigma}

\newcommand{\C}{\mathbb{C}}

\newcommand{\hlf}{\frac{1}{2}}

\newcommand{\non}{\nonumber}

\newcommand{\p}{\partial}
\renewcommand{\P}{\mathbb{P}}

\newcommand{\R}{\mathbb{R}}
\newcommand{\rr}{\rightarrow}
\newcommand{\w}{\wedge}
\newcommand{\Z}{\mathbb{Z}}
\newcommand{\GL}{\operatorname{GL}}
\renewcommand{\Im}{\operatorname{Im}}
\renewcommand{\O}{\operatorname{O}}
\renewcommand{\Re}{\operatorname{Re}}
\newcommand{\rk}{\operatorname{rk}}
\newcommand{\SL}{\operatorname{SL}}
\newcommand{\SO}{\operatorname{SO}}
\newcommand{\Sp}{\operatorname{Sp}}
\newcommand{\SU}{\operatorname{SU}}
\newcommand{\U}{\operatorname{U}}
\newcommand{\lp}{\left(}
\newcommand{\rp}{\right)}
\newcommand{\ls}{\left[}
\newcommand{\rs}{\right]}

\long\def\del#1\enddel{}        \long\def\new#1\endnew{{\bf #1}}

\def\ifundefined#1{\expandafter\ifx\csname#1\endcsname\relax}
\ifundefined{draftmode} {} \else        % \let\cite=\NEWcite
   \newcount\HOUR  \HOUR=\the\time \divide\HOUR by 60    \multiply \HOUR by 60
   \newcount\MIN   \MIN=\the\time  \advance\MIN by -\HOUR   \divide\HOUR by 60
   \ifnum   \MIN>9  \def\printTIME{{\it\the\HOUR\,:\,\the\MIN}}
         \else   \def\printTIME{{\it\the\HOUR\,:\,0\the\MIN}} \fi % \printTIME
   \renewcommand{\thepage}{\hbox to9pt{\hss{draft version~ \it \draftmode
        } ~ ~ --- ~ ~ } {\bf\arabic{page}}
        \hbox to9pt{ ~ ~ --- ~ ~ {generated \it\today} ~ ~at~ \printTIME\hss}}
\fi

\begin{document}
\preprint{
UTTG-02-07 \\
%{\tt hep-th/yymmnnn}\\
}
\title{Toroidal Orientifolds in IIA with General NS-NS Fluxes}

\author{ {\sc Matthias Ihl}, {\sc Daniel Robbins} and {\sc Timm Wrase}
     \oneaddress{
      Theory Group, Department of Physics,\\
      University of Texas at Austin,\\
      Austin, TX 78712, USA \\
      {~}\\
      \email{msihl@zippy.ph.utexas.edu}\\
      \email{robbins@zippy.ph.utexas.edu}\\
      \email{wrase@zippy.ph.utexas.edu}\\
      }
}

\date{May 23, 2007}

\Abstract{Type IIA toroidal orientifolds offer a promising toolkit for model builders, especially when one includes not only the usual fluxes from NS-NS and R-R field strengths, but also fluxes that are T-dual to the NS-NS three-form flux.  These new ingredients are known as metric fluxes and non-geometric fluxes, and can help stabilize moduli or can lead to other new features.  In this paper we study two approaches to these constructions, by effective field theory or by toroidal fibers twisted over a toroidal base.  Each approach leads us to important observations, in particular the presence of D-terms in the four-dimensional effective potential in some cases, and a more subtle treatment of the quantization of the general NS-NS fluxes.  Though our methods are general, we illustrate each approach on the example of an orientifold of $T^6/\Z_4$.}

\maketitle
\tableofcontents
\newpage
%%%%%%%%%%%%%%%%%%%%%%%%%%%%%%%%%%%%%%%%%%%%%%%%%%%%%%%%%%%%%%%%%%
\section{Introduction}

Much recent interest and research activity has been devoted to understanding the space of string theory vacua, especially those which can be described using the formalism of four-dimensional $\mathcal{N}=1$ supergravity.  As different constructions and compactifications have been explored, the number of tools in the model builder's kit has grown, even as our understanding of how they can be used and combined has sometimes diluted.  For instance string theory admits field strengths of various cohomological degree, and turning on fluxes of these field strengths through compact cycles of the internal space can often help stabilize the moduli of the compactification.  These include R-R fluxes and fluxes of the NS-NS three form field strength $H_3$.  This situation is fairly well understood.  Under T-duality, $H$-flux can sometimes be converted into twists of the internal space metric, which are known as geometric fluxes.  Further T-dualities can introduce so-called non-geometric fluxes, which ruin the global geometric description of the internal space, but still seem to give a consistent picture in the four-dimensional effective theory.  In fact, in the effective theory, one can in principle combine all of these fluxes, up to certain constraints and consistency conditions, but it has not been demonstrated that a ten-dimensional construction can necessarily always be found.

Our goal in this paper is to carefully explore how all of these ingredients can be combined in the context of $\mathcal{N}=1$ toroidal orientifolds of type IIA, though we believe that many of our methods can be applied in broader contexts.  To this end we will follow two different approaches, examining these constructions from the effective field theory point of view and also trying to present honest ten-dimensional constructions of as broad a class as possible.  Throughout the paper we will illustrate each method by referring to the example of an orientifold of $T^6/\Z_4$, whose structure is rich enough to illustrate many of the phenomena and techniques that we will describe.

In the effective field theory approach our primary goal is to classify the possible (untwisted sector) fluxes and translate them into the 4D $\mathcal{N}=1$ language.  We will find that the general NS-NS fluxes are most naturally parametrized by their action on the untwisted cohomology, along the lines described in~\cite{Tomasiello:2005bp,Aldazabal:2006up,Micu:2007rd,Cvetic:2007ju}.  So just as one can replace a discussion of the individual components $H_{ijk}$ of $H$-flux with coefficients $p_K$ in the expansion $H_3=p_Kb_K$, where $b_K$ are the untwisted three-forms which are anti-invariant under the orientifold action, one can also replace metric flux components $\om^i_{jk}$ by coefficients $r_{aK}$ and $\widehat r_{\al K}$, where $K$ again runs over three forms, and $a$ ($\al$) runs over invariant two-forms which are odd (even) under the orientifold involution.  Similarly, the nongeometric flux components $Q^{ij}_k$ and $R^{ijk}$ can be replaced by $q_{aK}$, $\widehat q_{\al K}$, and $s_K$.  In terms of these parameters it is then straight-forward to describe the data of the four-dimensional theory, and in particular we find the K\"ahler potential, the superpotential, and the holomorphic gauge couplings and D-terms.  There are additional consistency constraints that such general fluxes must satisfy; one of these is the R-R tadpole condition (to which the orientifold six-planes contribute), and there are also Bianchi identities, which are a set of constraints, quadratic in the NS-NS fluxes.  The tadpole condition can be elegantly expressed in terms of our cohomological flux parameters, but unfortunately the Bianchi identities only seem to be cleanly expressed using the original flux components.  In any given example, however, we may certainly express the Bianchi identities in our cohomological parameters, but the structure seems complicated and ad-hoc.

The presence of D-terms arising from general NS-NS fluxes is a phenomenon that has not, to our knowledge, been previously discussed in the literature.  We describe how adding certain metric fluxes (which are never simply T-duals of $H$-flux) can lead to electric charges for some of the four-dimensional scalar fields.  It will also turn out that certain non-geometric fluxes correspond to magnetic charges for the same fields, making them electric-magnetic dyons in general.  However, making use of the Bianchi identities one can show that the dyonic charges are necessarily mutually local, and there is always a consistent Lagrangian description of the effective theory.

As we introduce more general types of fluxes into our story, we will see how they enter in the particular example of $T^6/\Z_4$ with a certain orientifold action, and in particular we will look for supersymmetric solutions with as many moduli as possible stabilized.  For some simple cases, such as having only $H$-flux, or including certain classes of metric fluxes, we are able to find all supersymmetric solutions, but are unable to stabilize all moduli in these contexts.  For generic fluxes, subject to a naive quantization condition, we are able to numerically find supersymmetric solutions with all moduli stabilized.  Unfortunately, we will later learn that the naive quantization condition was, in fact, naive.  Using the correct quantization we can still stabilize all moduli, but are unable to satisfy the tadpole condition.  It seems likely, however, that this is not a result of a fundamental obstacle, but simply relates to a lack of understanding of the correct quantization of R-R fluxes in the presence of general NS-NS fluxes (or at least from not using the correct representatives for the K-theory or integral cohomology when using the twisted torus language).  We will also prove that fully stabilized supersymmetric Minkowski vacua (as opposed to AdS) require us to at least turn on non-geometric fluxes.

After exhausting ourselves in the playground of effective field theory, we then attempt to directly construct as many of these models as possible starting from ten dimensions, and following the approach of~\cite{Dabholkar:2002sy}.  We do this by splitting our $T^6$ into a base and a fiber, and then allowing the fiber to vary over the base.  The NS-NS fluxes are then encoded as twists of the fiber theory as we go around closed, non-contractible loops in the base.  We outline how to classify such splittings and twists for a given orientifold action, we show that consistency of our picture implies the Bianchi identities, and we also see clearly how to determine the correct quantization conditions on the NS-NS fluxes.  Simple integral quantization of the flux components or cohomological parameters turns out to be correct only in a sub-class of cases (which of course includes all situations with only $H$-flux, and all cases T-dual to those ones).

These constructions enjoy certain advantages; the action of the T-duality group is quite transparent.  This approach should easily generalize to many other interesting situations where a well-understood fiber theory is twisted over a toroidal base.  Of particular interest, we note that the flux combinations which occur in the low-energy effective theory are naturally described (as noted in~\cite{Shelton:2006fd,Micu:2007rd}) as a sort of covariant derivative on the spin-bundle whose sections are R-R-fields (see also the interesting discussion in~\cite{Ellwood:2006ya}).

In the context of our specific example, we will classify all base-fiber splittings and all fluxes which can be obtained in these constructions.  We will find that all $H$-fluxes and almost all metric fluxes can be turned on and a sub-class of non-geometric $Q$-fluxes can also be turned on.  Among the metric flux configurations that we can build are some which are not T-duals of $H$-flux alone, and in particular we can turn on D-terms and cases with non-standard quantization.  We cannot turn on any $R$-flux, which is not surprising, since there are arguments~\cite{Shelton:2006fd} that any construction giving rise to $R$-flux cannot have even a locally geometric description in ten dimensions.

The plan of this paper is as follows.  In section~\ref{BasicExample} we lay out our conventions for the $T^6/\Z_4$ orientifold which will be our canonical example throughout.  In section~\ref{EFTApproach} we delve into effective field theory, starting by describing in general our $\mathcal{N}=1$ language in section~\ref{N1Language}.  In section~\ref{OnlyHFlux} we add $H$-flux, first in the general story and then for our example.  Next, in section~\ref{AddingMetricFluxes} we include metric fluxes into the story, discussing the general framework in section~\ref{omegaGeneralFramework}, the D-terms in section~\ref{D-terms}, the induced superpotential in section~\ref{omegaSuperpotential} and the context of our example in section~\ref{omegaExample}.  Section~\ref{GeneralNSNSFluxes} introduces the non-geometric fluxes, and we revisit the D-terms in section~\ref{DTermsRevisited} and our example in~\ref{NonGeometricExample}.  In~\ref{Puzzles} we summarize this approach.

Then we turn to our base-fiber constructions in section~\ref{BFApproach}.  We introduce the T-duality group in section~\ref{TheTDualityGroup} and particularly how to discuss the orientifold action in the language of $\O(6,6)$.  In~\ref{NSNSFluxes} we describe how to encode the NS-NS fluxes as twists of our fibers, starting with a particular example for illustration before moving on to the general case.  Section~\ref{BFExample} is devoted to exploring these techniques in the $T^6/\Z_4$ example, including a complete classification of all twists possible with these constructions.  Some discussion is presented in~\ref{AdvantagesAndPuzzles}.

Finally, two appendices provide some extra detail.  Appendix~\ref{SU3Structure} compares our results with the literature on $\SU(3)$-structure and torsion cycles, providing a nice check on our formulae, as well as a purely geometric interpretation of the D-term constraints (they are equivalent to demanding that the manifold be half-flat).  And appendix~\ref{BIDerivation} provides two different derivations of the Bianchi identities, using the Jacobi identity for a certain Lie algebra, or alternatively by demanding that the covariant derivative, which encodes the action of the fluxes on the spin-bundle of the R-R-fields, squares to zero.

In the interests of carefully illustrating our techniques (and exploring them ourselves) we return repeatedly to the $T^6/\Z_4$ example in this paper, trying to push the ideas as far as possible in this specific context.  Unfortunately, the level of detail necessary in these sections is well beyond what is needed for a basic explanation of our results.  Readers only interested in the results and techniques should feel encouraged to skip any section or subsection with the word ``example" in the title, namely section \ref{BasicExample} and sections~\ref{HExample}, \ref{omegaExample}, \ref{NonGeometricExample}, \ref{TDualityGroupExample}, \ref{NSNSFluxExample}, \ref{BFExample}.  The other sections should be self-consistent.

\section{The Basic Example of $T^6/\Z_4$}
\label{BasicExample}

Our primary example throughout this paper will be a particular toroidal orientifold described below.  Before we dive into detailing this example and our conventions, the reader may be interested to know why we focus on this compactification, rather than one of the other orientifolds in the literature, such as $T^6/\Z_2^2$ or $T^6/\Z_3^2$ which have been more extensively studied and which are in some sense simpler.  We certainly believe that our approach here can be applied to these models.  One reason for our choice is familiarity, as two of the authors have studied this example in the past~\cite{Ihl:2006pp} and we are able to build on the solutions found there using T-duality.

However a more important reason to look at this example is that it, unlike the two other examples mentioned above, admits untwisted two-forms which are {\it even} under the orientifold involution.  By reducing the R-R potential $C_3$ along these forms we find four-dimensional vectors with associated $\U(1)$ gauge groups.  Later, in section~\ref{D-terms}, we will see that with certain metric fluxes turned on, some moduli become charged under the $\U(1)$s, and this gives rise to D-terms in the four-dimensional effective potential, a possibility that has not, to our knowledge, been discussed in the context of these models.

One final interesting property of this particular example that we do not make use of in the present work, is the existence of twisted sector three-forms, which again do not occur in the more well-studied models.  In principle these could lead to interesting possibilities for metric and non-geometric twisted-sector fluxes, in the spirit of~\cite{Cvetic:2007ju}.

\subsection{Setup}

We take the model from~\cite{Ihl:2006pp,Blumenhagen:2002gw}, namely a certain orientifold of $(T^2)^3$, but several of our conventions will differ, so we review everything here.  Let $z_1=x_1+iy_1$, $z_2=x_2+iy_2$, and $z_3=x_3+(\hlf+iU)y_3$ be complex coordinates on the tori (we will see below that $U$ is a real modulus parametrizing the complex structure of the third torus), and the torus identifications are given by integer shifts in each $x_i$ or $y_i$.  The orientifold group is generated by a $\Z_4$ rotation
\be
\Theta:\lp z_1,z_2,z_3\rp\longrightarrow\lp iz_1,iz_2,-z_3\rp,
\ee
and the orientifold action is $\Om_p(-1)^{F_L}\s$, where the antiholomorphic involution $\s$ acts as
\be
\s:\lp z_1,z_2,z_3\rp\longrightarrow\lp\bar z_1,i\bar z_2,\bar z_3\rp.
\ee
Note that $\Theta\s=\s\Theta^3$, so the full orientifold group is in fact isomorphic to the dihedral group $D_4$.  This model is frequently referred to as an orientifold of the orbifold $T^6/\Z_4$ even though it is not a $\Z_2$ quotient of the orbifold.  Rather, the precise statement is that the full orientifold group is a $\Z_2$ extension of the $\Z_4$ orbifold group.  We emphasize this point now partly as a warning to the reader, since we will likely be guilty of sloppy language at times in the work below.

This orientifold is the {\bf ABB} model in the classification of~\cite{Blumenhagen:1999ev}.

\subsection{Cohomology}

We will begin by describing the untwisted cohomology of $T^6/\Z_4$, dividing further into subspaces which are even or odd under the involution $\s$.  The bases we will present will consist of elements of $H^\ast(T^6;\Z)$ with the correct symmetry properties.  In this way we get bases for the untwisted cohomology of the orbifold {\it over the rationals}.  The correct quantization conditions for fluxes in the orientifold are subtle, and should in principle require an understanding of the correct K-theory analog for our model which would go beyond the scope of this paper~\cite{DFM}.  Instead we will point out where such information would be relevant, and explain why we do not believe that it will affect our results significantly.

We start with the even cohomology, implicitly equating classes with their harmonic form representatives.  There is one zero form, namely the unit function $1$.  For two-forms, there are five independent $(1,1)$-forms invariant under the rotations: four odd forms,
\bea
\om_1 &=& \frac{i}{2}dz_1\w d\bar z_1=dx_1\w dy_1,\non\\
\om_2 &=& \frac{i}{2}dz_2\w d\bar z_2=dx_2\w dy_2,\\
\om_3 &=& \frac{i}{2U}dz_3\w d\bar z_3=dx_3\w dy_3,\non\\
\om_4 &=& \frac{1-i}{2}\lp dz_1\w d\bar z_2-idz_2\w d\bar z_1\rp\non\\
&=& dx_1\w dx_2-dx_1\w dy_2+dy_1\w dx_2+dy_1\w dy_2,\non
\eea
and one even form
\bea
\m &=& \frac{1+i}{2}\lp dz_1\w d\bar z_2+idz_2\w d\bar z_1\rp\non\\
&=& dx_1\w dx_2+dx_1\w dy_2-dy_1\w dx_2+dy_1\w dy_2.
\eea

Similarly, for four-forms we have four even $(2,2)$-forms
\be
\widetilde\om_1=\om_2\w\om_3,\qquad\widetilde\om_2=\om_1\w\om_3,\qquad\widetilde\om_3=\om_1\w\om_2,\qquad\widetilde\om_4=\om_3\w\om_4,
\ee
and one odd $(2,2)$-form,
\be
\widetilde\m=\om_3\w\m.
\ee

Finally there is one six-form, which is odd under the involution,
\be
\varphi=\om_1\w\om_2\w\om_3=dx_1\w dy_1\w dx_2\w dy_2\w dx_3\w dy_3.
\ee

The nonzero integrals involving these forms over $X=T^6/\Z_4$ are (wedge products are implicit)
\be
\int_X\varphi=\int_X\om_1\,\om_2\,\om_3=\frac{1}{4},\qquad\int_X\om_3\,\om_4^2=\int_X\om_3\,\m^2=-1,\non
\ee
\be
\int_X\om_1\,\widetilde\om_1=\int_X\om_2\,\widetilde\om_2=\int_X\om_3\,\widetilde\om_3=\frac{1}{4},\qquad\int_X\om_4\,\widetilde\om_4=\int_X\m\,\widetilde\m=-1.
\ee

Next we have the odd cohomology.  It turns out that $H^1(X)$ and $H^5(X)$ are empty, so we need only describe the three-forms.  The basis we shall use is
\bea
a_1 &=& \chi_{xxx}+\chi_{xxy}+\chi_{xyx}+\chi_{yxx}-\chi_{yyx}-\chi_{yyy},\non\\
a_2 &=& \chi_{xxx}+\chi_{xyx}+\chi_{xyy}+\chi_{yxx}+\chi_{yxy}-\chi_{yyx},\\
b_1 &=& -\chi_{xxx}+\chi_{xyx}+\chi_{xyy}+\chi_{yxx}+\chi_{yxy}+\chi_{yyx},\non\\
b_2 &=& \chi_{xxx}+\chi_{xxy}-\chi_{xyx}-\chi_{yxx}-\chi_{yyx}-\chi_{yyy}.\non
\eea
Here we use notation where $\chi_{xyx}=dx_1\w dy_2\w dx_3$, etc.  The forms $a_I$ are even under the involution $\s$, while $b_I$ are odd.  The nonzero integrals are simply $\int_Xa_I\w b_J=\d_{IJ}$.

\subsection{Moduli}
\label{Moduli}

With the basis of differential forms given above, we can now describe the various moduli of this model.  Most of this work will focus entirely on the untwisted sector, so we shall start there, with a brief description of the twisted sectors at the end of the subsection.

Our choice of complex coordinates has already determined the $(3,0)$-form $\Om$ up to an overall constant factor.  We shall fix the phase of this factor by demanding that $\s\cdot\Om=\bar\Om$, and fix the modulus with the requirement that
\be
i\int_X\Om\w\bar\Om=1.
\ee
With these requirements, $\Om$ is determined up to an overall sign which we simply pick by hand, giving
\bea
\Om &=& \frac{1-i}{2\sqrt{U}}\,dz_1\w dz_2\w dz_3\\
&=& \frac{1}{2\sqrt{U}}\ls\lp\hlf+U\rp a_1+\lp\hlf-U\rp a_2+i\lp\hlf+U\rp b_1-i\lp\hlf-U\rp b_2\rs.
\eea
In this expression, $U$ is the unique untwisted complex structure modulus, and is a real variable in the range $0<U<\infty$.

For the K\"ahler form we can write
\be
J=v_a\,\om_a,
\ee
where $v_a$, $a=1,\ldots,4$, are the real K\"ahler moduli of the metric.  The corresponding line element is
\bea\label{metric}
ds^2 &=& v_1\lp dx_1^2+dy_1^2\rp+v_2\lp dx_2^2+dy_2^2\rp\non\\
&& +\frac{v_3}{U}\lp dx_3^2+dx_3dy_3+\lp\frac{1}{4}+U^2\rp dy_3^2\rp\\
&& -2v_4\lp dx_1dx_2+dx_1dy_2-dy_1dx_2+dy_1dy_2\rp.\non
\eea
In order for the metric to have the correct (euclidean) signature, we must have
\be
v_1>0,\qquad v_2>0,\qquad v_3>0,\qquad{\mathrm{and}}\qquad v_1v_2-2v_4^2>0.
\ee
The volume is given by
\be
{\mathcal V}_6=\frac{1}{3!}\int_XJ^3=\frac{1}{4}v_3\lp v_1v_2-2v_4^2\rp.
\ee
Note also that having $J$ odd under the anti-holomorphic involution $\s$ implies that the metric is invariant under $\s$, as required for the orientifold projection.  This is why there is no allowed metric deformation corresponding to the even two-form $\m$.

These moduli pair up with periods of the $B$-field (which must be odd under $\s$ to survive projection),
\be
B=u_a\,\om_a,
\ee
to give the complex K\"ahler moduli $t_a=u_a+iv_a$ and the corresponding complexified K\"ahler form,
\be
J_c=t_a\,\om_a=B+iJ.
\ee

The untwisted NS-NS sector moduli are then completed by adding in the dilaton $\phi$.  From the R-R sector, we have only periods of the three form $C_3$.  In order to survive the projection, this form must be even under the action of $\s$, so we have only two real moduli $\xi_I$, $I=1,2$, where
\be
C_3=\xi_I\,a_I.
\ee

Let us now quickly summarize the twisted sector moduli.  The fixed locus of $\Theta$ or $\Theta^3$ consists of sixteen points, eight of which are fixed by $\s$, plus four pairs of points that get swapped by $\s$.  The four pairs will give rise to four even and four odd $(1,1)$-forms and equal numbers of even and odd $(2,2)$-forms.  The remaining eight points will each contribute an odd $(1,1)$-form and an even $(2,2)$-form. 
%{\bf [I don't know how to prove this... presumably it would suffice to know the metric in complex coordinates for the blow-up of $\C^3/\Z_4$, which should be something like $\mathcal{O}_{\mathbb{F}_2}(-4,-2)$ over the Hirzebruch surface $\mathbb{F}_2$; see section A.2 of hep-th/0609040.]}  
All together, then, these twisted sectors contribute twelve new complex K\"ahler moduli, with four moduli of the orbifold being projected out by the orientifold.

The fixed locus of $\Theta^2$ consists of sixteen two-tori.  $\Theta$ invariance gives four copies of $T^2/\Z_2$, and six pairs of $T^2$ that get interchanged.  Each of these six pairs automatically contributes one even and one odd form.  Of the six pairs, two pairs are at $\s$ fixed points and each contributes an odd $(1,1)$-form and an even $(2,2)$-form\footnote{To see that a $\s$ fixed point gives rise to an odd $(1,1)$-form, note that locally it looks like $(\C^2/\Z_2)\times T^2$ which resolves to $(\mathcal{O}_{\P^1}(-2))\times T^2$.  An explicit K\"ahler metric can be written down for the latter geometry and one can check that the unique normalizable $(1,1)$-form is odd under an antiholomorphic involution.}, two more pairs have $\s$ act the same as $\Theta$ and hence also act as if they were $\s$ fixed points, and the final two pairs are interchanged by $\s$, leading to one odd and one even of each $(1,1)$- and $(2,2)$-forms.  Similarly the remaining four $T^2/\Z_2$ are each at fixed points of $\s$, and each contribute an odd $(1,1)$-form and an even $(2,2)$-form.  In total then, this sector contains ten two-forms, nine of which are odd, twelve three-forms, which split into six odd and six even, and ten four-forms, nine of which are even.

\subsection{Fluxes}
\label{Fluxes}

Finally, we turn to the allowed fluxes which we can turn on in our model.  As mentioned above, the correct classification of R-R fluxes in this model would involve a careful discussion of K-theory in this setting, and would go beyond the scope of this paper.  We will instead stick to cohomology.  Moreover, we will be primarily interested in so-called ``bulk fluxes" - fluxes whose image in rational cohomology has a nonzero projection onto the untwisted sector.  For this reason, we will write our fluxes as (recalling that $F_0$ and $F_4$ need to be even under $\s$, while $F_2$ and $F_6$ need to be odd)
\bea
F_0 &=& m_0,\non\\
F_2 &=& m_a\,\om_a,\\
F_4 &=& e_a\,\widetilde\om_a,\non\\
F_6 &=& e_0\varphi,\non
\eea
where quantization conditions say that\footnote{Note the unusual factor of $\sqrt 2$ in this expression.  This is a consequence of the form of the R-R kinetic terms in our conventions.  See also the discussion in~\cite{DeWolfe:2005uu}.}
\be
\frac{\sqrt{2}}{(2\pi)^{p-1}(\al')^{(p-1)/2}}\int\,F_p\in\Z,
\ee
with the integral taken over any $p$-cycle in $X$.  This is of course not completely correct; proper quantization requires combining untwisted and twisted sector fluxes, but it is not quite as bad as one might fear.  Indeed, one can argue (see e.g.~\cite{Frey:2002hf,DeWolfe:2005uu}) that any bulk flux can be written as one of the above, plus twisted sector contributions which correspond to fractional fluxes at fixed points (modulo again certain K-theoretic subtleties).

The NS-NS three-form flux is in some sense simpler.  It must simply lie in $H^3(X;\Z)\cap H_{\mathrm{odd}}^\ast(X)$.  In principle this can be completely worked out - the difficult step of working out the integral cohomology has already been done in~\cite{Blumenhagen:2002gw} - but again we won't need the full detail and we shall again project onto the untwisted sector cohomology.  This allows us to write
\be
H_3=p_I\,b_I,
\ee
and impose the simple quantization $p_I/(4\pi^2\al')\in\Z$.  We will also set $\al'=1/4\pi^2$ unless otherwise noted.

At any rate, we will not make too much use of the underlying quantizations of the R-R fluxes (though we will encounter a puzzle related to this later on), and the quantization of NS-NS fluxes will be treated much more carefully in section~\ref{BFApproach}.

\subsection{Orientifold planes}

Orientifold planes will lie at the fixed locus of each orientation-reversing element of the orientifold group.  For instance, the fixed locus of the involution $\s$ is described by the set
\be
\lp T^6\rp^\s=\left\{0\le x_1\le 1,y_1\in\{0,\hlf\};0\le x_2=y_2\le 1;0\le x_3\le 1,y_3=0\right\}\subset T^6.
\ee
In homology, this cycle can be written as
\be
\ls\lp T^6\rp^\s\rs=2\pi_{xxx}+2\pi_{xyx},
\ee
where $\pi_{xxx}$ is the cycle represented by $y_i$ fixed, $x_i$ variable and winding once.  We also have, e.g.,
\be
\int_{\pi_{xxx}\subset T^6}\chi_{xxx}=1,
\ee
with other choices of integrand giving vanishing results.  Using this, we have also picked the orientation of this cycle to be such that it is positively calibrated by $\operatorname{Re}\Om$.

Similar consideration for the other anti-holomorphic involutions of the orientifold group give
\bea
\ls\lp T^6\rp^{\Theta\s}\rs &=& 2\pi_{xyx}-4\pi_{xyy}+2\pi_{yyx}-4\pi_{yyy},\non\\
\ls\lp T^6\rp^{\Theta^2\s}\rs &=& 2\pi_{yxx}-2\pi_{yyx},\\
\ls\lp T^6\rp^{\Theta^3\s}\rs &=& -2\pi_{xxx}+4\pi_{xxy}+2\pi_{yxx}-4\pi_{yxy}.\non
\eea
The total class of the O6-plane is thus
\be
\ls O6\rs=4\lp\pi_{xxy}+\pi_{xyx}-\pi_{xyy}+\pi_{yxx}-\pi_{yxy}-\pi_{yyy}\rp.
\ee
It is then easy to verify that
\be
\int_{\ls O6\rs\subset X}a_1=4,\qquad\int_{\ls O6\rs\subset X}a_2=\int_{\ls O6\rs\subset X}b_1=\int_{\ls O6\rs\subset X}b_2=0,
\ee
where we must be careful to divide by four relative to the result on $T_6$ (this doesn't make sense for individual cycles like $\pi_{xxx}$ which are not invariant under the orbifold action, but does make sense for the total class $\ls O6\rs$).

The reason that it is important to know where the orientifold plane lies is that it can contribute to the tadpole for the seven-form R-R potential, $C_7$.  Explicitly, the equation of motion for $C_7$ includes a contribution proportional to $\d_{\mathrm{O6}}$, the delta-three form supported on the orientifold plane.  From the computations above, we see that in cohomology,
\be
[\d_{\mathrm{O6}}]=4b_1.
\ee

\section{Effective Field Theory Approach}
\label{EFTApproach}

Much of the work that has been done on toroidal orientifolds with fluxes turned on has been in the context of a four-dimensional effective field theory description.  This is not at all surprising; in backgrounds with R-R fluxes alone there is a lack of satisfactory world-sheet descriptions.  Similarly, NS-NS fluxes can be tricky to deal with in a world-sheet formalism, and in some of the more exotic cases (such as $R$-fluxes~\cite{Shelton:2005cf}, which will be discussed below) no ten-dimensional description is known.

So the approach has been to start with ten-dimensional situations in which there is a description and reduce to a four-dimensional effective theory.  We will be working in $\mathcal{N}=1$ language, and so starting from a given set of fluxes, we will need to give a four-dimensional superpotential, a K\"ahler potential, and holomorphic gauge couplings.  From this data we obtain our effective theory.  Using the T-duality group, which is also a duality group of the low-energy theory, people have been able to guess the four-dimensional data for more general sets of NS-NS fluxes~\cite{Aldazabal:2006up,Micu:2007rd}.

In the sections below we will go over these arguments for increasingly more general sets of fluxes, giving the general result and then applying it to our specific example.

\subsection{$\mathcal{N}=1$ language}
\label{N1Language}

Before we turn on any fluxes, it is useful to review the $\mathcal{N}=1$ supergravity theory we will be constructing in four dimensions and see what we can learn already.  Such a theory  will generally consist of one gravity multiplet, some number of chiral multiplets including complex scalars $\phi^I$ and some number of vector multiplets including vectors $A^\al$.  The theory is then specified by giving three functions which will depend on the complex scalars, namely a K\"ahler potential $K$, a holomorphic superpotential $W$, and holomorphic gauge-kinetic couplings $f_{\al\beta}$.  The bosonic part of the effective action is then
\bea
S^{(4)} &=& -\int_{M^4}\left\{-\hlf R\ast 1+K_{I\bar J}d\phi^I\w\ast d\bar\phi^{\bar J}+V\ast 1\right.\non\\
&& \left.+\hlf\lp\Re f_{\al\beta}\rp F^\al\w\ast F^\beta+\hlf\lp\Im f_{\al\beta}\rp F^\al\w F^\beta\right\},
\label{N=1Action}
\eea
where the scalar potential is
\be
V=e^K\lp K^{I\bar J}D_IW\,D_{\bar J}\bar W-3|W|^2\rp+\hlf\lp\Re f\rp^{-1\,\al\beta}D_\al D_\beta.
\label{N=1ScalarPotential}
\ee
Here, $\ast$ is the four-dimensional Hodge star, $K_{I\bar J}=\p_I\bar\p_{\bar J}K$, $K^{I\bar J}$ is its inverse, $F^\al=dA^\al$, and $D_IW=\p_IW+(\p_IK)W$.  $D_\al$ is the D-term for the $\U(1)$ gauge group corresponding to $A^\al$, i.e.
\be
\label{Dalpha}
D_\al=\p_IK\,\lp T_\al\rp^I\vphantom{\lp T^\al\rp}_J\phi^J+\zeta_\al,
\ee
where $T_\al$ is the generator of the gauge group, and $\zeta_\al$ is the Fayet-Iliopoulos term.

Let us now consider how this effective theory is obtained, following~\cite{Grimm:2004ua}, from the ten-dimensional models in which we are interested.  We will take the situation (as in our example) where besides the constant zero-form $1$ and the volume form $\phi$, we have odd two-forms $\om_a$, even two-forms $\m_\al$, even three-forms $a_K$, odd three-forms $b_K$, even four-forms $\widetilde\om_a$, and odd four-forms $\widetilde\m_\al$, where the index $a$ runs from $1$ to $h^{1,1}_{-\,(\mathrm{untwisted})}$, $\al$ runs from $1$ to $h^{1,1}_{+\,(\mathrm{untwisted})}$, and $K$ runs from $1$ to $h^{2,1}_{(\mathrm{untwisted})}+1$.  The intersection numbers are taken to be
\be
\int_X\varphi=\frac{1}{|\G|},\qquad\int_X\om_a\,\om_b\,\om_c=\k_{abc},\qquad\int_X\om_a\widetilde\om_b=d_{ab},
\ee
\be
\int_X\m_\al\,\m_\beta\,\om_a=\widehat\k_{\al\beta a},\qquad\int_X\m_\al\widetilde\m_\beta=\widehat d_{\al\beta},\qquad\int_Xa_J\w b_K=\d_{JK},\non
\ee
where wedge products are implicit between even forms and where $|\G|$ is the order of the orbifold group (four, for our example).

The four-dimensional chiral fields will be related to moduli of the ten-dimensional theory.  First there are the K\"ahler moduli, $t_a=u_a+iv_a$, from
\be
B+iJ=J_c=t_a\om_a,
\ee
(the complexified K\"ahler form $J_c$ should be an odd two-form).  To describe the complex moduli, let us write the holomorphic three-form as
\be
\Om=\mathcal{Z}_Ka_K-\mathcal{F}_Kb_K,
\ee
and we take the conventions (as in section~\ref{Moduli}) that
\be
i\int_X\Om\w\bar\Om=1,\qquad\s\cdot\Om=\bar\Om,
\ee
so that the $\mathcal{Z}_K$ are real functions of the complex structure moduli of the metric, while the $\mathcal{F}_K$ are pure imaginary, and together they satisfy the constraint $\mathcal{Z}_K\mathcal{F}_K=-i/2$.  For comparison with section~\ref{Moduli}, we have in our example
\be
\mathcal{Z}_1=\frac{1}{2\sqrt U}\lp\hlf+U\rp,\qquad\mathcal{Z}_2=\frac{1}{2\sqrt U}\lp\hlf-U\rp,\non
\ee
\be
\mathcal{F}_1=\frac{-i}{2\sqrt U}\lp\hlf+U\rp,\qquad\mathcal{F}_2=\frac{i}{2\sqrt U}\lp\hlf-U\rp.
\ee
We then define a complexified version
\be
\Om_c=C_3+2ie^{-D}\Re\Om=\lp\xi_K+2ie^{-D}\mathcal{Z}_K\rp a_K,
\ee
where $e^{-D}=\mathcal{V}_6^{1/2}e^{-\phi}$ contains the dilaton, and $\mathcal{V}_6=\frac{1}{6}\k_{abc}v_av_bv_c$ is the volume.  The complex moduli $N_K=\hlf\xi_K+ie^{-D}\mathcal{Z}_K$ are simply given by the expansion
\be
\Om_c=2N_Ka_K.
\ee

Similarly, the four-dimensional vectors will come from reducing $C_3$ against the forms $\m^\al$, so that the total field $C_3$ (before turning on fluxes), is
\be
C_3=\xi_Ka_K+A^\al\w\m_\al.
\ee

We would next like to derive the functions $K$, $W$, and $f_{\al\beta}$ by reducing the ten-dimensional action for type IIA, which in the Einstein frame reads
\bea
S^{(10)} &=& -\hlf\int_{M^4\times X}\left\{-R\ast 1+\hlf d\phi\w\ast d\phi+\hlf e^{-\phi}H_3\w\ast H_3+e^{\frac{5}{2}\phi}F_0\ast F_0+e^{\frac{3}{2}\phi}F_2\w\ast F_2\right.\non\\
&& \left.+e^{\hlf\phi}F_4\w\ast F_4+B_2\w dC_3\w dC_3+B_2^2\w dC_3\w dC_1+\frac{1}{3}B_2^3\w dC_1\w dC_1\right.\non\\
&& \left.+\frac{1}{3}F_0B_2^3\w dC_3+\frac{1}{4}F_0B_2^4\w dC_1+\frac{1}{20}F_0^2B_2^5\right\},
\label{MassiveIIAAction}
\eea
where in the absence of fluxes,
\be
\label{NoFluxes}
B_2=u_a\om_a,\qquad H_3=du_a\w\om_a,\qquad F_0=0,\qquad C_1=0,\qquad F_2=dC_1+F_0B_2=0,
\ee
\be
C_3=\xi_Ka_K+A^\al\w\m_\al,\qquad F_4=dC_3+C_1\w H_3+\hlf F_0B_2^2=d\xi_K\w a_K+F^\al\w\m_\al.\non
\ee
Plugging these into (\ref{MassiveIIAAction}) and then integrating over $X$ we can compare the resulting four-dimensional action with (\ref{N=1Action}).  For example, comparing the coeffecient of $F^\al\w F^\beta$ we find that
\be
\Im f_{\al\beta}=u_a\int_X\om_a\w\m_\al\w\m_\beta=\widehat\k_{\al\beta\,a}u_a,
\ee
and the coefficient of $F^\al\w\ast F^\beta$ gives
\be
\Re f_{\al\beta}=e^{\hlf\phi}\int_X\m_\al\w\ast\m_\beta=-\widehat\k_{\al\beta\,a}v_a,
\ee
where we have converted to string frame and used an expression for $\int_X\m_\al\w\ast\m_\beta$ in terms of intersection numbers, as found e.g. in~\cite{Grimm:2004ua}.  Alternatively, we could have just used $\Im f_{\al\beta}$ and our knowledge of the holomorphicity of $f_{\al\beta}$.  Either way we conclude
\be
f_{\al\beta}=i\widehat\k_{\al\beta\,a}t_a.
\ee

Similarly (though with more effort) we find that $W=0$ and the K\"ahler potential is given by
\be
K=4D-\ln\lp 8\mathcal{V}_6\rp=4D-\ln\lp\frac{4}{3}\k_{abc}v_av_bv_c\rp,
\ee
and this expression should of course be thought of as a real function of the complex fields $t_a$ and $N_K$, defined implicitly through its dependence on $v_a$ and $D$.

We will see that the effect of turning on fluxes will be to introduce a nonzero superpotential $W$, but that the kinetic terms for the four-dimensional fields will not be affected, and hence neither $f_{\al\beta}$ nor $K$ will change.

Finally, one finds that the D-term contribution to the scalar potential also vanishes, and we conclude that all the FI parameters are vanishing and that all complex scalars are neutral under each gauge group.  It will be useful for later contexts to think briefly about how one checks the neutrality under gauge transformations here.  Note that a gauge transformation
\be
A^\al\rightarrow A^\al+d\lambda^\al,
\ee
is inherited from the ten-dimensional gauge transformation
\be
C_3\rightarrow C_3+d\lp\lambda^\al\m_\al\rp,
\ee
which preserves the form of the expansion of $C_3$.  Under this transformation, $A^\al$ is the only field which changes.  In sections below, we will find that the two-forms $\m_\al$ are no longer necessarily closed, and so the ten-dimensional gauge transformation above will also be felt by some of the scalar fields.  This effect will be interpreted as a charge on a given field, and so in this case D-terms will be generated (though we will find that the FI parameters will continue to vanish in our setup).

\subsection{Including only $H$-flux}
\label{OnlyHFlux}

To begin, we start by turning on arbitrary R-R fluxes and only the most familiar sort of NS-NS flux, namely $H$-flux.  This situation has been studied extensively, and we primarily follow the work of~\cite{Grimm:2004ua,DeWolfe:2005uu}, (though the conventions we will use differ slightly, and are engineered to agree with~\cite{Aldazabal:2006up,Shelton:2005cf,Shelton:2006fd}).

\subsubsection{General results}

As in section~\ref{Fluxes}, we expand our fluxes in our cohomological basis as
\be
F_0=m_0,\qquad F_2=m_a\om_a,\qquad F_4=e_a\widetilde\om_a,\qquad F_6=e_0\varphi,\ee
and
\be
H_3=p_Kb_K.
\ee
These are in addition to the contributions from the moduli as seen in (\ref{NoFluxes}).  As mentioned above, $K$ and $f_{\al\beta}$ are given as before, and the gauge transformation argument proceeds unchanged, showing that there are no charged scalars.  Finally, performing an explicit reduction shows that the FI parameters continue to vanish (so there are no D-terms), and there is now a superpotential given by
\be
W=W^Q+W^K,
\ee
with
\be
\label{HW}
W^Q=\int_X\Om_c\w H_3,\qquad W^K=\int_X e^{J_c}\w F_{\mathrm{RR}},
\ee
where
\be
e^{J_c}=1+J_c+\hlf J_c\w J_c+\frac{1}{6}J_c\w J_c\w J_c,
\ee
and
\be
F_{\mathrm{RR}}=F_0+F_2+F_4+F_6,
\ee
are formal sums of forms.  Performing the integrals over $X$, we find
\be
W=2N_Kp_K+\frac{1}{|\G|}e_0+d_{ab}t_ae_b+\hlf\k_{abc}t_at_bm_c+\frac{m_0}{6}\k_{abc}t_at_bt_c.
\label{HSuperpotential}
\ee

Another very important point which we have ignored up to now is the presence of a tadpole for the R-R field $C_7$.  This field is nondynamical, explaining why we have not included it above, but its tadpole must nonetheless be cancelled.  Indeed, the ten-dimensional action has a piece
\be
\label{C7Tadpole}
\int_{M^4\times X}\left\{-\hlf\lp F_2+m_0B\rp\w\ast\lp F_2+m_0B\rp+C_7\w\ls\frac{1}{\sqrt 2}\d_{D6}-\sqrt 2\d_{O6}\rs\right\},
\ee
where the $\d$s are delta-function three forms representing the localized sources.  Since $\ast(F_2+m_0B)=\widetilde F_8=dC_7+\cdots$, the $C_7$ equation of motion then implies that
\be
-m_0p_Kb_K+\frac{1}{\sqrt 2}\ls\d_{D6}\rs=\sqrt 2\ls\d_{O6}\rs,
\ee
(though note that the tadpole condition is actually stronger than this cohomological version).  Because of the freedom to use D-branes (or anti-D-branes if necessary) to satisfy the tadpole condition, we will attempt first to find vacua without worrying about the tadpole condition, and then see what, if anything, we then need to add.

To find supersymmetric vacua, we need to solve the F-term equations $D_aW=0$ and $D_KW=0$.  From the results above, and using the useful fact that
\be
\label{pKK}
\p_KD=-e^D\mathcal{F}_K,
\ee
we find the real and imaginary parts of these equations to be
\bea
0 &=& \Re D_aW=d_{ab}e_b+\k_{abc}u_bm_c+\frac{m_0}{2}\k_{abc}\lp u_bu_c-v_bv_c\rp-\frac{3}{2}\frac{\k_{abc}v_bv_c}{\k_{def}v_dv_ev_f}\Im W,\label{HReDaW}\\
0 &=& \Im D_aW=\k_{abc}v_bm_c+m_0\k_{abc}u_bv_c+\frac{3}{2}\frac{\k_{abc}v_bv_c}{\k_{def}v_dv_ev_f}\Re W,\label{HImDaW}\\
0 &=& \Re D_KW=2p_K-4ie^D\mathcal{F}_K\Im W,\label{HReDKW}\\
0 &=& \Im D_KW=4ie^D\mathcal{F}_K\Re W.
\label{HImDKW}
\eea

Since not all of the $\mathcal{F}_K$ can vanish (recall the condition $i\int_X\Om\w\bar\Om=1$), (\ref{HImDKW}) requires $\Re W=0$.  One can quickly check that a Minkowski solution (one in which $\Im W$ also vanishes) will force all of the fluxes (R-R and NS-NS) to vanish; in this case the superpotential vanishes, no moduli are stabilized, and the tadpole must be saturated by adding D6-branes.

Suppose now that we are not in a Minkowski, but rather an AdS solution, in which $\Im W\ne 0$.  In order for the metric to be positive definite, the matrix $(\k v)_{ab}=\k_{abc}v_c$ should be invertible, and so equation (\ref{HImDaW}) tells us that either $m_0=0$ and $m_a=0$, or
\be
u_a=-\frac{m_a}{m_0}.
\ee
The former case reduces to the unstabilized Minkowski vacuum mentioned above.

Also, if the F-term equations hold, then one can subtract $e^{-D}\mathcal{Z}_K\Re D_KW+v_a\Re D_aW$ from the imaginary part of the right hand side of (\ref{HSuperpotential}) to show that
\be
\Im W=-\frac{2m_0}{15}\k_{abc}v_av_bv_c.
\label{HImW}
\ee

One can proceed somewhat further in the general case, but since we would like to add more ingredients to our construction, we will refer the reader to~\cite{DeWolfe:2005uu}, and restrict ourselves instead to our specific example.

\subsubsection{Example}
\label{HExample}

Now focus on our $T^6/\Z_4$ orientifold, assuming an AdS solution to the F-term equations.  Then since we must have $\mathcal{F}_1\ne 0$ for a nondegenerate solution, equations (\ref{HReDKW}) tell us that $p_1\ne 0$ and that
\be
\label{HU}
\frac{\mathcal{F}_2}{\mathcal{F}_1}=\frac{p_2}{p_1}\qquad\Longrightarrow\qquad U=\hlf\frac{p_1+p_2}{p_1-p_2},
\ee
From this we see that a sensible solution requires $|p_1|>|p_2|$.

Next, we use (\ref{HReDaW}) and (\ref{HImW}) to obtain a set of four quadratic equations for the $v_a$.  The equations are simplest if we write them in terms of quantities
\be
\widehat e_1=e_1-\frac{m_2m_3}{m_0},\qquad\widehat e_2=e_2-\frac{m_1m_3}{m_0},\qquad\widehat e_3=e_3-\frac{m_1m_2-2m_4^2}{m_0},\qquad\widehat e_4=e_4-\frac{m_3m_4}{m_0}.
\ee
It turns out that a sensible solution (i.e. one in which $v_1$, $v_2$, $v_3$ are all positive and $v_1v_2>2v_4^2$) exists if and only if $m_0$, $\widehat e_1$, $\widehat e_2$, $\widehat e_3$ are all the same sign and if $\widehat e_1\widehat e_2>2\widehat e_4^2$.  If these conditions are met, then we have a sensible, physical solution given by
\be
v_1=\left|\widehat e_2\right|\sqrt{\frac{5}{3m_0}\frac{\widehat e_3}{\widehat e_1\widehat e_2-2\widehat e_4^2}},\qquad v_2=\left|\widehat e_1\right|\sqrt{\frac{5}{3m_0}\frac{\widehat e_3}{\widehat e_1\widehat e_2-2\widehat e_4^2}},
\ee
\be
v_3=\sqrt{\frac{5}{3m_0}\frac{\widehat e_1\widehat e_2-2\widehat e_4^2}{\widehat e_3}},\qquad v_4=\widehat e_4\sqrt{\frac{5}{3m_0}\frac{\widehat e_3}{\widehat e_1\widehat e_2-2\widehat e_4^2}}\lp\operatorname{sign} m_0\rp.\non
\ee

Next we can solve for the dilaton.  It turns out that $e^D>0$ implies that $p_1$ must have the opposite sign of $m_0$, and then
\be
e^D=\ls\frac{27m_0}{10}\frac{p_1^2-p_2^2}{\widehat e_3\lp\widehat e_1\widehat e_2-2\widehat e_4^2\rp}\rs^{1/2},\qquad\mathrm{or}\qquad e^\phi=\frac{3}{2}\sqrt{p_1^2-p_2^2}\ls\frac{12}{5}m_0\widehat e_3\lp\widehat e_1\widehat e_2-2\widehat e_4^2\rp\rs^{-1/4}.
\ee

Finally, we can use $\Re W=0$ to solve for one linear combination of the axions, $p_K\xi_K$; the scalar potential is independent of the other linear combination and so this other combination remains a flat direction perturbatively.
\be
p_1\xi_1+p_2\xi_2=-\frac{1}{4}e_0+\frac{1}{4m_0}\lp m_1e_1+m_2e_2+m_3e_3-4m_4e_4\rp-\frac{m_3\lp m_1m_2-2m_4^2\rp}{2m_0^2}.
\ee

Thus, for a given general set of fluxes (satisfying certain inequalities) we have found the unique solution to the F-term equations and have found that all but one of the moduli are fixed.  We still, however, need to satisfy the tadpole constraint.  Indeed, since $[\d_{O6}]=4b_1$, we find
\be
-\sqrt 2m_0p_1+N_1=8,\qquad-\sqrt 2m_0p_2+N_2=0,
\ee
where $N_1$ and $N_2$ are the number of D-branes wrapping the cycle dual to $b_1$ or $b_2$ respectively.  Actually one needs to be a bit careful here; a supersymmetric D-brane should have a positive volume as calibrated by $\Re\Om$, but for the cycle dual to $b_2$ the orientation picked out by this condition depends on whether $U$ is less than or greater than one half.  If $U<\hlf$, as is the case when $m_0p_2>0$, then we should have $N_2>0$ D6 branes, in agreement with the above.  On the other hand, if $U>\hlf$, then the cycle dual to $b_2$ is negatively calibrated and $N_2$ counts the number of anti-D6 branes.  In this case $N_2<0$ for a SUSY solution, but we also have $m_0p_2<0$, so the tadpole condition can still be satisfied.

Note that to find a physical solution above, we required that $m_0p_1<0$, and hence immediately $N_1<8$.  In fact, since we also needed $|p_1|\ge|p_2|$, we have that $N_1+|N_2|<8$; the total number of D6-branes is bounded.  Hence, we see that in some sense the fluxes here contribute to the tadpole with the same sign as the D-branes.  We are not allowed to add as many D-branes as we like to saturate the tadpole, but rather (within SUSY) our gauge groups have bounded rank.

Before moving on, let us note that we could have worked directly with the scalar potential of (\ref{N=1ScalarPotential}) and looked for extrema of the potential.  Recall that if we have an extremum at which the value of the potential is negative, so that we have AdS$_4$, then stability does not require that the the extremum be an actual minimum.  It is enough that each field $\Phi^I$ have a mass squared that is greater than the Breitenlohner-Freedman bound,
\be
m_I^2>-\frac{3}{4}\left|V_{\mathrm{extremum}}\right|,
\ee
where we assume that $\Phi^I$ has a canonically normalized kinetic term.  Indeed, for the supersymmetric solution above, it turns out that there is one mode with a negative mass squared, but it is above the BF bound.

\subsection{Adding metric fluxes}
\label{AddingMetricFluxes}

The next ingredient we will be adding is known as {\it{metric flux}}.  It is well known that by T-dualizing one circle of a torus with $H$-flux, one can swap the $H$-flux for some nonconstant metric components.  One finds that some of the original globally defined one-forms of the torus, $dx^i$, are no longer globally defined, but need to be replaced by a set of one-forms $\eta^i$,\footnote{To be precise, the space of globally defined smooth one-forms on the torus is spanned by the $dx^i$ with coefficients that are smooth, globally defined functions on the torus (e.g. $1$, $\cos x$, $2\sin 5x$, etc.).  Similarly on the twisted torus the smooth, globally defined one-forms are spanned by the $\eta^i$ with smooth global functions as coefficients.  In fact, the entire ring $\Lambda^\bullet T^\ast(X)$ is generated from the $\eta^i$ over smooth functions in this way.} which are no longer necessarily closed (see~\cite{Kachru:2002sk,Marchesano:2006ns} for a discussion of the cohomology of twisted tori), but rather satisfy
\be
\label{MaurerCartan}
d\eta^i=-\hlf\om^i_{jk}\eta^j\w\eta^k,
\ee
where $\om^i_{jk}$ are constant coefficients, antisymmetric in the lower two indices.  These coeffecients are known as metric (or sometimes geometric) fluxes, and arise from the NS-NS sector of the theory, just as the $H$-flux does.

In fact, one needs not necessarily obtain such solutions by T-duality, but rather one can start from (\ref{MaurerCartan}) directly.  An effective four-dimensional theory can still be obtained by performing a generalized Scherk-Schwarz reduction.  This has been done in some other models in~\cite{Villadoro:2005cu,Shelton:2005cf,Shelton:2006fd}, and some general work has also been done~\cite{Scherk:1979zr,Kaloper:1998kr,Kaloper:1999yr}, but we will point out a couple of novel features, such as the appearance of nonvanishing D-terms, which have not been explored in these models before.

There are some subtle issues here about the general consistency of this program which we will discuss more in sections~\ref{Puzzles} and~\ref{BFApproach}, but for now we shall forge ahead.

\subsubsection{General framework}
\label{omegaGeneralFramework}

First, let us note that by taking the exterior derivative of (\ref{MaurerCartan}), we find that $d^2=0$ provides a consistency condition,
\be
\label{eq:BI1}
\omega^m_{[ij}\omega^n_{k]m}=0, \; \forall n,i,j,k,
\ee
We will refer to this condition, along with similar conditions for the other fluxes, as Bianchi identities which the NS-NS fluxes will need to satisfy.

Another perspective on these fluxes which is sometimes useful is that if we write
\be
\eta^i=N^i_j(x)dx^j,
\ee
then we can construct vector fields
\be
Z_i=\lp N^{-1}\rp^j_i\frac{\p}{\p x^j}.
\ee
These turn out to be Killing vectors of the twisted torus, and they form a Lie algebra,
\be
\ls Z_i,Z_j\rs=\om^k_{ij}Z_k.
\ee
The Jacobi identity for the Lie algebra simply reproduces (\ref{eq:BI1}).  This algebra is somewhat useful to keep in mind and we can sometimes relate properties of the system with metric fluxes to properties of the algebra.  We will discuss these matters in sections~\ref{Puzzles} and~\ref{BFApproach}.

Another identity which must be satisfied is that the $H$-flux, which we will now write as\footnote{Note that it is important that $H_3$ be a globally defined three-form.}
\be
H_3=H_{ijk}\eta^i\w\eta^j\w\eta^k,
\ee
must still be closed, leading to the Bianchi identity
\be
\om^i_{[jk}H_{\ell m]i}=0.
\ee
There is one more constraint that we will impose, namely that traces $\om^i_{ij}=0$ for all $j$.  One can obtain this constraint for instance by demanding that the volume form of the torus not be exact, since
\be
d\lp\frac{1}{5!}\e_{ij_1\cdots j_5}\eta^{j_1}\w\cdots\w\eta^{j_5}\rp=\om^j_{ji}\eta^1\w\cdots\w\eta^6.
\ee
Happily, this condition will be automatically true anytime we are in a space $X$ with $H^1(X)=0$; the orbifold projection will not allow any object with a single free index.  All of these Bianchi identities will in some sense be unified below in section~\ref{Shorthand} and in Appendix~\ref{BIDerivation}.

As just mentioned, since our interest here is in toroidal orientifolds of type IIA, we must restrict our choices of $\om^i_{jk}$ so that they are invariant under the full orientifold group.  We implicitly followed the same procedure for the components $H_{ijk}$ of $H$-flux, only there we required that they be invariant under the orbifold group and odd under the involution, since the worldsheet parity operator $\Om$ which accompanies the involution flips the sign of $B_2$.  For the metric fluxes the story is similar, except that since the metric is even under worldsheet parity, we find that $\om^i_{jk}$ should be even under the involution.  This story can be told more cleanly in the base-fiber approach in section~\ref{BFApproach}.

In the case of the $H_{ijk}$, it was then natural to parametrize our choices of flux not by the individual components that remained after projection, but by the coefficients $p_K$ in the expansion $H_3=p_Kb_K$.  A similar choice can be made for the metric fluxes which will vastly simplify our discussion of the effective action and the tadpole constraint.  To this end, consider a general $p$-form,
\be
A^{(p)}=\frac{1}{p!}A_{i_1\cdots i_p}\eta^{i_1}\w\cdots\w\eta^{i_p},
\ee
where we will assume here that the coefficients $A_{i_1\cdots i_p}$ are constants.  Then we will define a $(p+1)$-form $\om\cdot A=-dA$, which in components as above reads\footnote{The reason for introducing this notation here rather than simply using the exterior derivative is so that we can more easily unify the results with nongeometric fluxes introduced below.}
\be
\lp\om\cdot A\rp_{i_1\cdots i_{p+1}}=\binom{p+1}{2}\om^j_{[i_1i_2}A_{|j|i_3\cdots i_{p+1}]},
\ee
and where we are using conventions such that $\binom{n}{m}=0$ unless $0\le m\le n$.

Then we can now define coefficients $r_{aK}$ and $\widehat r_{\al K}$ from the expansions
\be
\om\cdot\om_a=r_{aK}b_K,\qquad\om\cdot\m_\al=\widehat r_{\al K}a_K.
\ee
Integration by parts then also furnishes the expansions
\be
\om\cdot a_K=\lp d^{-1}\rp^{ab}r_{bK}\widetilde\om_a,\qquad\om\cdot b_K=-\lp{\widehat d}^{-1}\rp^{\al\beta}\widehat r_{\beta K}\widetilde\m_\al.
\ee
Note that we are abusing notation here slightly, since we are using the same symbols to denote our forms, which are now expanded in the $\eta$ basis, e.g. $\om_1=\eta^{x_1}\w\eta^{y_1}$.  We will expand our fluxes in this new basis of forms as before, using integers $m_0$, $m_a$, $e_a$, and $e_0$ for $F_0$, $F_2$, $F_4$, and $F_6$.  The lack of invariant one- and five-forms ensures that the fluxes associated to $F_0$, $F_4$, $F_6$ remain closed, and the Bianchi identity ensures that $H_3$ is closed, but we now see that the flux $m_a\om_a$ corresponding to $F_2$ is not.

Indeed, by looking at (\ref{C7Tadpole}), we see that there will be a new contribution to the $C_7$ tadpole,
\be
-\sqrt 2\lp m_0p_K-m_ar_{aK}\rp b_K+\ls\d\rs_{D6}=2\ls\d\rs_{O6}.
\ee

Actually, it is worthwhile to briefly rephrase some of these results with an eye toward later sections.  As has long been noted in more general geometric setups with $H$-flux, such as this one, it can be useful to define a modified formal derivative
\be
d_H=d+H\w=H\w\cdot-\om\cdot,
\ee
The requirement that $H$ be closed can be obtained from imposing $d_H^2=0$, and in our case this recovers all of our quadratic Bianchi identities.  Many of the Bianchi identities can be found by applying $d_H^2$ to our cohomological basis,
\bea
d_H^2\,1=-p_K\lp{\widehat d}^{-1}\rp^{\al\beta}\widehat r_{\beta K}\widetilde\m_\al &\Longrightarrow& p_K\widehat r_{\al K}=0,\,\forall\al,\label{dH2}\\
d_H^2\,\om_a=r_{aK}\lp{\widehat d}^{-1}\rp^{\al\beta}\widehat r_{\beta K}\widetilde\m_\al &\Longrightarrow& r_{aK}\widehat r_{\beta K}=0,\,\forall a,\beta,\non
\eea
but sadly, as we shall see in our specific example, this does not capture all of the Bianchi identities.
And finally, the contribution to the tadpole is naturally proportional to
\be
-\left.d_HF_{RR}\right|_{\mathrm{3-form}}=-\lp HF_0-\om\cdot F_2\rp,
\ee
in precise agreement with what we have found.  

\subsubsection{D-terms}
\label{D-terms}

Now let us revisit the gauge transformations for the four-dimensional vectors $A_\al$.  Recall that the vectors descended from the three-form potential,
\be
C_3=A^\al\w\m_\al+\xi_Ka_K,
\ee
where we ignore the local piece of $C_3$ which contributes to the four-form flux $e_a\widetilde\om_a$.  In order to generate the required gauge transformation $A^\al\rightarrow A^\al+d\lambda^\al$, we perform a gauge transformation
\be
C_3\longrightarrow C_3+d\lp\lambda^\al\m_\al\rp=C_3+d\lambda^\al\w\m_\al-\lambda^\al\widehat r_{\al K}a_K,
\ee
or in terms of the four dimensional fields,
\be
A^\al\longrightarrow A^\al+d\lambda^\al,\qquad \xi_K\longrightarrow\xi_K-\lambda^\al\widehat r_{\al K},
\ee
We thus see that our scalar fields are no longer all invariant under the gauge transformations!  In particular, if we define a field
\be
\label{XiK}
\Xi_K=\exp\ls iN_K\rs,
\ee
then $\Xi_K$ is electrically charged under the gauge group $\U(1)_\al$ with charge $-\hlf\widehat r_{\al K}$.  Using (\ref{pKK}) and (\ref{Dalpha}), we can then calculate
\be
D_\al=-2ie^D\mathcal{F}_K\widehat r_{\al K},
\ee
(recall that our $\mathcal{F}_K$'s were pure imaginary, so that $D_\al$ is real).  So for a supersymmetric vacuum we must have, in addition to the F-term equations that we will derive below, that $\mathcal{F}_K\widehat r_{\al K}=0$ for each gauge group $\al$.  Note that no Fayet-Iliopoulos terms have been generated.

On the other hand, if we are willing to break SUSY, we note that the scalar potential now has a piece
\be
\label{VD}
V_D=\hlf\lp\Re f\rp^{-1\,\al\beta}D_\al D_\beta=-2e^{2D}\lp\Re f\rp^{-1\,\al\beta}\lp\mathcal{F}_K\widehat r_{\al K}\rp\lp\mathcal{F}_J\widehat r_{\beta J}\rp,
\ee
where we recall that
\be
\Re f_{\al\beta}=-\widehat\k_{\al\beta a}v_a.
\ee
Since $(\Re f)$ is positive definite, $V_D\ge 0$.  In appendix~\ref{SU3Structure} we give a geometric interpretation for the nonvanishing of the D-terms.

Such D-term contributions have been the subject of much phenomenological interest, as a possible means to uplift the potential to a metastable deSitter vacuum~\cite{Burgess:2003ic,Villadoro:2005yq,Achucarro:2006zf,Parameswaran:2006jh,Parameswaran:2007kf}, or as a mechanism for generating inflationary potentials~\cite{Binetruy:1996xj,Halyo:1996pp}.  The latter possibility is usually done in a context with FI parameters turned on, but with a minimal holomorphic coupling $f_{\al\beta}=\d_{\al\beta}$; it would be interesting to see if the class of models we are discussing in this paper could lead to phenomenologically useful potentials.  We are currently investigating these possibilities.

\subsubsection{Superpotential}
\label{omegaSuperpotential}

One can obtain the superpotential in the presence of metric fluxes in a number of ways.  One can perform an explicit generalized Scherk-Schwarz reduction, or the formula can be deduced from T-duality arguments, as has been done by various authors~\cite{Shelton:2005cf,Shelton:2006fd,Banks:2006hg,Aldazabal:2006up}.  Using the formalism of generalized complex geometry is also an interesting approach (see~\cite{Benmachiche:2006df,Micu:2007rd}).

The result is that the superpotential can still be written as $W=W^Q+W^K$, with $W^K=\int\exp[J_c]\w F_{RR}$, exactly as before, but with $W^Q$ now modified into
\be
\label{omegaWQ}
W^Q=\int_X\Om_c\w\lp H_3+\om\cdot J_c\rp=\int_X\Om_c\w d_H\lp e^{-J_c}\rp,
\ee
where $\Om_c$ and $J_c$ are as before, so that doing the integration,
\be
W^Q=2N_K\lp p_K+r_{aK}t_a\rp.
\ee
In particular, $W^Q$ no longer depends only on the complex moduli, but there is now a mixing term $2N_Kr_{aK}t_a$.

Given that we discovered in section~\ref{D-terms} above that some of our scalar fields now transform under the gauge groups, an immediate worry is whether the superpotential is neutral, as it must be for consistency.  Computing, we find
\be
\d W=-\lambda^\al\widehat r_{\al K}\lp p_K+r_{aK}t_a\rp=0,
\ee
where in the final step we have used the Bianchi identities (\ref{dH2}) that arise from applying $d_H^2$ to our cohomological basis.  So our setup seems consistent.

We turn now to the F-term equations that result from this superpotential.
\bea
0 &=& \Re D_aW=r_{aK}\xi_K+d_{ab}e_b+\k_{abc}u_bm_c+\frac{m_0}{2}\k_{abc}\lp u_bu_c-v_bv_c\rp-\frac{3}{2}\frac{\k_{abc}v_bv_c}{\k_{def}v_dv_ev_f}\Im W,\non\\
&&\label{omegaReDaW}\\
0 &=& \Im D_aW=2e^{-D}r_{aK}\mathcal{Z}_K+\k_{abc}v_bm_c+m_0\k_{abc}u_bv_c+\frac{3}{2}\frac{\k_{abc}v_bv_c}{\k_{def}v_dv_ev_f}\Re W,\label{omegaImDaW}\\
0 &=& \Re D_KW=2p_K+2r_{aK}u_a-4ie^D\mathcal{F}_K\Im W,\label{omegaReDKW}\\
0 &=& \Im D_KW=2r_{aK}v_a+4ie^D\mathcal{F}_K\Re W.\label{omegaImDKW}
\eea

It is once again true in this case that one can use the F-term equations to show
\be
\Im W=-\frac{2m_0}{15}\k_{abc}v_av_bv_c.
\ee
Thus, if we would like to find a supersymmetric Minkowski vacuum, we must have $m_0=0$.  In such a vacuum, (\ref{omegaImDKW}) says that $r_{aK}v_a=0$, and then contracting (\ref{omegaImDaW}) with $v_a$ we learn that $\k_{abc}v_av_bm_c=0$.  With a couple more manipulations one can then show that the F-term equations now reduce to
\be
M\cdot\lp\begin{matrix}u_a \\ \xi_K\end{matrix}\rp+\lp\begin{matrix}\lp de\rp_a \\ p_K\end{matrix}\rp=0,\qquad M\cdot\lp\begin{matrix}\widetilde v_a \\ 2\mathcal{Z}_K\end{matrix}\rp=0,
\ee
where
\be
M=\lp\begin{matrix}\lp\k m\rp_{ab} & r_{aK} \\ r^T_{Jb} & 0\end{matrix}\rp,
\ee
and where $\widetilde v_a=e^Dv_a$.  Two more equations must also be satisfied - there is one relation among the $\mathcal{Z}_K$ ($\mathcal{Z}_1^2-\mathcal{Z}_2^2=\hlf$ in our example), and from $\Re W=0$ we have
\be
\frac{1}{|\G|}e_0+d_{ab}u_ae_b+\hlf\k_{abc}u_au_bm_c=0.
\ee
Since the $v_a$ and $D$ only occur in the combination $\widetilde v_a$, there will always be one combination which remains unfixed (this result was also derived by~\cite{Micu:2007rd}).  Explicitly, the mode which scales $e^\phi=g_s\rightarrow\lambda g_s$ and $v_a\rightarrow\lambda^2v_a$ leaves $\widetilde v_a$ unchanged, so this mode will remain massless.  The scaling here is unfortunate; it means that as we go far out along this flat direction, either the string coupling blows up or the volume becomes very small, and our whole framework is expected to break down.  This means that we cannot expect parametric control of such an example.

The general situation for supersymmetric AdS vacua, or even more generally for extrema of the full scalar potential, is quite complicated, and we don't have much to say about it here.  We do believe that with metric fluxes it should be possible to stabilize all moduli in a supersymmetric AdS vacuum, though a puzzle regarding R-R quantization will interfere with our attempts to provide a fully consistent example here.  

Let us now examine the situation in our specific example more closely.

\subsubsection{Example}
\label{omegaExample}

Imposing invariance under the orientifold group, we find that we are left with ten independent metric fluxes,
\bea
2 & \omega^1_{16} &= \omega^1_{15} = -\omega^2_{25} =-2 \, \omega^2_{26},\non\\
&\omega^1_{26} &= \omega^2_{16},\non\\
&\omega^1_{36} &=-\omega^2_{46},\non\\
&\omega^1_{46} &=\omega^2_{36},\non\\
&& \quad \omega^1_{35}=\omega^1_{45}=\omega^2_{35} = -\omega^2_{45}=\omega^1_{36}+\omega^1_{46},\non\\
& \omega^3_{16} &= -\omega^4_{26},\non\\
& \omega^3_{26} &= \omega^4_{16},\\
&& \quad \omega^3_{15} =\omega^3_{25}=\omega^4_{15}=-\omega^4_{25}=\omega^3_{16} +\omega^3_{26},\non\\
& \omega^3_{36} &=- \omega^4_{46},\non\\
2 &\omega^3_{46} &=\omega^3_{45}=\omega^4_{35}=2 \, \omega^4_{36},\non\\
& \omega^5_{13} &=-\omega^5_{24},\non\\
& \omega^6_{13} &=-\omega^6_{14}=-\omega^6_{23}=-\omega^6_{24},\non\\
&& \quad \omega^5_{14}=\omega^5_{23}=\omega^5_{13}+\omega^6_{13},\non
\eea
where we can use the ten fluxes in the left-hand column as representatives, and where (here and elsewhere in the paper) we order our coordinates $(x_1,y_1,x_2,y_2,x_3,y_3)$, so that an index $2i-1$ refers to $x_i$ and an index $2j$ refers to $y_j$.

In terms of $r$-matrices, we find
\be
\label{rMatrixExample}
r_{aK}=\lp\begin{matrix}\om^1_{36} & -\om^1_{46} \\ -\om^3_{16} & \om^3_{26} \\ \om^5_{13}+\om^6_{13} & \om^5_{13} \\ \om^1_{16}-\om^1_{26}-\om^3_{36}-\om^3_{46} & -\om^1_{16}-\om^1_{26}-\om^3_{36}+\om^3_{46}\end{matrix}\rp,
\ee
\be
\label{widehatrMatrixExample}
\widehat r_K=\lp\begin{matrix}-\om^1_{16}+\om^1_{26}-\om^3_{36}-\om^3_{46} & \  & -\om^1_{16}-\om^1_{26}+\om^3_{36}-\om^3_{46}\end{matrix}\rp.
\ee
Note that there is a one-to-one correspondence between the independent fluxes $\om^i_{jk}$ and the entries of $r$ and $\widehat r$.

Let us now impose the Bianchi identities $\om^m_{[jk}\om^i_{\ell]m}=0$ and $\om^m_{[ij}H_{kl]m}=0$.  It turns out that the general solution can be divided into four cases.
\be
\begin{array}{clll}(a) & r=\lp\begin{matrix}0 & 0 \\ 0 & 0 \\ \al & \beta \\ 0 & 0\end{matrix}\rp, & \widehat r=0, & \forall p_1,p_2 \\ (a') & r=\lp\begin{matrix}0 & 0 \\ 0 & 0 \\ \al & -\al \\ 0 & 0\end{matrix}\rp, & \widehat r=\lp\begin{matrix}\beta & \beta\end{matrix}\rp, & p_1+p_2=0, \\ (a'') & r=\lp\begin{matrix}0 & 0 \\ 0 & 0 \\ \al & \al \\ 0 & 0\end{matrix}\rp, & \widehat r=\lp\begin{matrix}\beta & -\beta\end{matrix}\rp, & p_1-p_2=0, \\ (b) & r=\lp\begin{matrix}\al & \beta \\ \g & \d \\ 0 & 0 \\ \varepsilon & \varphi\end{matrix}\rp, & \widehat r=\lp\begin{matrix}\chi & \k\end{matrix}\rp, & \chi p_1+\k p_2=0,\end{array}
\ee
and where case $(b)$ must additionally satisfy the equations
\be
\al\chi+\beta\k=\g\chi+\d\k=\varepsilon\chi+\varphi\k=0,\qquad 8\al\g-\varepsilon^2-\chi^2=8\beta\d-\varphi^2-\k^2.
\ee
The first set of these equations is simply $r_{aK}\widehat r_K=0$, as we derived above in (\ref{dH2}).  The one remaining equation, however, cannot be obtained from acting $d_H^2$ on any element of our orbifold-invariant cohomology, though it can be derived by demanding $d_H^2=0$ even on non-invariant forms.

Let us try to find supersymmetric solutions to these models.  First of all, note that in cases $(a')$ and $(a'')$ the D-term is proportional to
\be
\beta\lp\mathcal{F}_1\pm\mathcal{F}_2\rp,
\ee
which is always nonvanishing since $|\mathcal{F}_1|>|\mathcal{F}_2|$ in nondegenerate ($0<U<\infty$) vacua.  Hence, these two cases can never be supersymmetric.  So we shall instead examine case $(a)$ more carefully.  Here the D-term equations are automatically satisfied since $\widehat r=0$.

Now using (\ref{omegaImDKW}), and assuming that at least one of $\al$ and $\beta$ is nonzero, we find that $\al\mathcal{F}_2=\beta\mathcal{F}_1$, so
\be
U=\hlf\frac{\al+\beta}{\al-\beta}.
\ee
For a physical solution, we need $|\al|>|\beta|$.  Then from (\ref{omegaReDKW}) we learn that in order to solve the F-term equations we have an extra condition on the fluxes, namely that
\be
\label{alphap2}
\al p_2=\beta p_1.
\ee
There are a couple of immediate consequences of this.  Firstly, observe that if $p_1\ne 0$, then $U$ actually has the same form (\ref{HU}) as before, and we again have that $|p_1|>|p_2|$.  Also, note that the axions $\xi_1$ and $\xi_2$ appear in the F-term equations only in the combinations $r_{aK}\xi_K$ and $p_K\xi_K$, but thanks to (\ref{alphap2}), both of these are proportional to $(\al\xi_1+\beta\xi_2)$; the equations are independent of the other linear combination, and hence one of the axions remains unfixed.  Below, we will argue that this will happen generically if the rank of the matrix $r_{aK}$ (one, for case $(a)$) is less than the number of axions in the problem (two).

We can now express the general solution to the F-term equations.  First we define some useful quantities,
\be
\widehat e_1=e_1-\frac{m_2m_3}{m_0},\qquad\widehat e_2=e_2-\frac{m_1m_3}{m_0},\qquad\widehat e_3=e_3-\frac{m_1m_2-2m_4^2}{m_0},\non
\ee
\be
\widehat e_4=e_4-\frac{m_3m_4}{m_0},\qquad\widehat e_0=e_0-\frac{e_1e_2-2e_4^2}{m_3}.
\ee
Then we find that in addition to $U$ given above, we have
\bea
u_1 &=& -\frac{m_1}{m_0}-\frac{\al\widehat e_2}{\al m_3-p_1m_0},\non\\
u_2 &=& -\frac{m_2}{m_0}-\frac{\al\widehat e_1}{\al m_3-p_1m_0},\\
u_3 &=& -\frac{m_3}{m_0}+\frac{5\al}{m_0}\frac{\lp\widehat e_1\widehat e_2-2\widehat e_4^2\rp\lp\al m_3-p_1m_0\rp}{3\lp\al m_3-p_1m_0\rp\lp\al\widehat e_0-p_1\widehat e_3\rp+\al\lp 5\al-3p_1\frac{m_0}{m_3}\rp\lp\widehat e_1\widehat e_2-2\widehat e_4^2\rp},\non\\
u_4 &=& -\frac{m_4}{m_0}-\frac{\al\widehat e_4}{\al m_3-p_1m_0}.\non
\eea
In terms of $u_3$ above, we then have
\be
v_3=\sqrt{-\frac{1}{\al m_0}\lp m_3+m_0u_3\rp\lp p_1+\al u_3\rp},
\ee
and
\bea
\al\xi_1+\beta\xi_2 &=& -\frac{\al}{4}\ls\frac{m_3\widehat e_3-m_0\widehat e_0}{\al m_3-p_1m_0}+\frac{m_0\lp\al m_3+p_1m_0\rp\lp\widehat e_1\widehat e_2-2\widehat e_4^2\rp}{m_3\lp\al m_3-p_1m_0\rp^2}\rs,\non\\
v_1 &=& -\frac{5}{3v_3}\frac{p_1+\al u_3}{\al m_3-p_1m_0}\widehat e_2,\\
v_2 &=& -\frac{5}{3v_3}\frac{p_1+\al u_3}{\al m_3-p_1m_0}\widehat e_1,\non\\
v_4 &=& -\frac{5}{3v_3}\frac{p_1+\al u_3}{\al m_3-p_1m_0}\widehat e_4,\non\\
e^\phi &=& \frac{3\sqrt{\al^2-\beta^2}\lp\al m_3-p_1m_0\rp}{2\sqrt 2\left|\al m_0\right|}\sqrt\frac{v_3}{\widehat e_1\widehat e_2-2\widehat e_4^2}.
\eea
A good solution requires a number of inequalities and conditions to hold; $\al m_3>p_1m_0$, $\widehat e_1\widehat e_2>2\widehat e_4^2$, and $(\al m_3-p_1m_0)(\al \widehat e_0-p_1\widehat e_3)>\al p_1m_0(\widehat e_1\widehat e_1-2\widehat e_4^2)/m_3$, and the quantities $\widehat e_1$, $\widehat e_2$, and $m_0$ must have the same sign.

As long as these conditions are respected, we can take various limits of the above solution.  For instance, one can check that taking the limit $\al, \beta\rr 0$ (and using (\ref{alphap2})) recovers the solution from section~\ref{HExample}.  For future reference, let us list also the limit $p_1,p_2\rr 0$.  In this case the conditions are that $\al$, $m_3$, $\widehat e_0$ must have the same sign, $\widehat e_1$, $\widehat e_2$, and $m_0$ must have the same sign, and $\widehat e_1\widehat e_2>2\widehat e_4^2$.
\be
u_1=-\frac{e_2}{m_3},\qquad u_2=-\frac{e_1}{m_3},\qquad u_3=-\frac{3m_3^2\widehat e_0}{m_0\lp 3m_3\widehat e_0+5\lp\widehat e_1\widehat e_2-2\widehat e_4^2\rp\rp},\qquad u_4=-\frac{e_4}{m_3},\non
\ee
\be
v_1=\left|\widehat e_2\right|\sqrt\frac{5\widehat e_0}{3m_3\lp\widehat e_1\widehat e_2-2\widehat e_4^2\rp},\qquad v_2=\left|\widehat e_1\right|\sqrt\frac{5\widehat e_0}{3m_3\lp\widehat e_1\widehat e_2-2\widehat e_4^2\rp},\non
\ee
\be
\label{NoHCaseaSolution}
v_3=\frac{\sqrt{\frac{15m_3^3}{m_0^2}\widehat e_0\lp\widehat e_1\widehat e_2-2\widehat e_4^2\rp}}{3m_3\widehat e_0+5\lp\widehat e_1\widehat e_2-2\widehat e_4^2\rp},\qquad  v_4=\widehat e_4\sqrt\frac{5\widehat e_0}{3m_3\lp\widehat e_1\widehat e_2-2\widehat e_4^2\rp}\lp\operatorname{sign}m_0\rp,
\ee
\be
\al\xi_1+\beta\xi_2=-\frac{1}{4}\ls\widehat e_3-\frac{m_0\widehat e_0}{m_3}+\frac{m_0}{m_3^2}\lp\widehat e_1\widehat e_2-2\widehat e_4^2\rp\rs,\non
\ee
\be
e^\phi=\frac{3\sqrt{\al^2-\beta^2}}{2\sqrt 2}\left|\frac{m_3}{m_0}\right|\frac{\ls\frac{15m_3^3}{m_0^2}\frac{\widehat e_0}{\widehat e_1\widehat e_2-2\widehat e_4^2}\rs^{1/4}}{\sqrt{3m_3\widehat e_0+5\lp\widehat e_1\widehat  e_2-2\widehat e_4^2\rp}}.\non
\ee
We will see later that these two limits are T-duals of each other.

Returning to the general case, note that the tadpole condition is now
\be
\sqrt 2\lp\al m_3-p_1m_0\rp+N_1=8,\qquad \sqrt 2\lp\beta m_3-p_2m_0\rp+N_2=0.
\ee
If we are looking for a supersymmetric solution, then we want $N_1$ to be greater than zero, and the sign of $N_2$ to be fixed by $U$ (as discussed in section~\ref{HExample}), and from solving the F-term equations we have $\al m_3>p_1m_0$, $\al p_2=\beta p_1$, and $|\al|>|\beta|$, so we find, as before, that $N_1+|N_2|<8$.

To help understand why we were unable to find solutions with all moduli stabilized in these examples, note that in the general situation the $h^{2,1}+1$ real fields $\xi_K$ appear in the F-term equations only through the combinations $p_K\xi_K$ and $r_{aK}\xi_K$.  We see immediately that these provide at most $\rk(r)+1$ independent combinations, so that we only have any chance to stabilize all of the axions in the case that $\rk(r)\ge h^{2,1}$ (see also the discussion in~\cite{Micu:2007rd}).  In fact, if $\Re W\ne 0$, then we can do even better.  In this case we can use (\ref{omegaImDKW}) to show that
\be
\mathcal{F}_K=\frac{i}{2}e^{-D}\frac{r_{aK}v_a}{\Re W},
\ee
and then (\ref{omegaReDKW}) implies that
\be
p_K=\lp-u_a-\frac{\Im W}{\Re W}v_a\rp r_{aK},
\ee
thus reducing us to just $\rk(r)$ independent combinations of axions.  In our example, this means that if $\Re W\ne 0$, we need an $r$-matrix of rank two, which was obviously impossible in the context of case $(a)$ above.  On the other hand, trying to set $\Re W=0$ seems to typically lead to degenerate solutions, where either the complex structure modulus or a K\"ahler modulus runs off to the edge of physically allowed values.

So finally, let us turn to case $(b)$, with the hopes of finding an $\mathcal{N}=1$ vacuum with all moduli fixed.  Suppose first that $\widehat r\ne 0$.  In this case, the D-term equations require $|\chi|<|\k|$ and fix $\mathcal{F}_2/\mathcal{F}_1=-\chi/\k$.  Then the various Bianchi identities enforce $r_{a2}=-(\chi/\k)r_{a1}$ and $p_2=-(\chi/\k)p_1$.  It is immediately clear in this case that one combination of the $\xi$'s again remains unfixed.

Hence, let $\widehat r=0$.  By the argument above, we should look for solutions in which $\rk(r)=2$.  As we will see in section~\ref{BFApproach}, the quantization conditions on metric fluxes is in general not the naive quantization in terms of (even) integers, but is somewhat more complicated.  In fact, we will see that we cannot find a correctly quantized set of metric fluxes which both give $\rk(r)=2$ and make it possible to satisfy the tadpole condition, however we think that this is a reflection of our ignorance of the correct R-R quantization conditions under these circumstances.  For now, let us willfully ignore these subtleties and pick a set of NS-NS fluxes with the naive quantization, namely
\be
p_1=p_2=0,\qquad r=\lp\begin{matrix}4 & 2 \\ 2 & 0 \\ 0 & 0 \\ 8 & 0\end{matrix}\rp.
\ee
This choice respects the Bianchi identities.  Let us then also choose R-R fluxes
\be
m_0=m_1=m_2=0,\qquad e_0=e_1=e_2=-e_3=e_4=\sqrt 2,\qquad -m_3=m_4=\frac{1}{\sqrt 2},
\ee
which all satisfy that they are in $\Z/\sqrt 2$.  One can check that these choices satisfy the tadpole conditions with no extra branes.  Then one can solve the F-term equations with these fluxes, first finding exact solutions for the $u_a$ and $\xi_K$
\be
u_1=0,\qquad u_2=2,\qquad u_3=-\frac{9}{2},\qquad u_4=-\hlf,\non
\ee
\be
\xi_1=-\frac{1}{4\sqrt 2},\qquad \xi_2=\frac{1}{2\sqrt 2},
\ee
and then solving numerically for the rest of the moduli,
\be
v_1\approx 2.58227,\qquad v_2\approx 4.26420,\qquad v_3\approx 3.46108,\qquad v_4\approx -0.50562,\non
\ee
\be
U\approx 1.03530,\qquad e^\phi\approx 11.956.
\ee
Note that the volumes here are not particularly large, though all the cycle volumes are positive, and the string coupling is definitely not small.  This solution should be viewed more as an in principle proof that the F-term equations can stabilize all of the moduli at physical values.

\subsection{General NS-NS fluxes}
\label{GeneralNSNSFluxes}

As we shall see when we consider T-dualities below, by T-dualizing twice on a torus with $H$-flux, one can find oneself in a {\it non-geometric} situation, where there is a local geometric decription, but globally, one must patch torus fibers together with non-geometric elements of the T-duality group.  All of this will hopefully be elucidated more cleanly in the next section, but for now note that at least some such considerations are really forced upon us by T-duality.  To this end we will introduce objects $Q^{ij}_k$, analogous to $H_{ijk}$ and $\om^i_{jk}$.

If one believes that the full $\O(6,6;\Z)$ T-duality group acts on the fluxes which appear in the effective four-dimensional description (this is not obviously correct from a ten-dimensional perspective; choosing a flux trivialization reduces the number of available isometries and correspondingly the size of the duality group), then one also should include fluxes which come from dualizing all three legs of $H$-flux as well, which will be denoted $R^{ijk}$.

\subsubsection{General approach}
\label{Shorthand}

As mentioned, we introduce fluxes $Q^{ij}_k$ and $R^{ijk}$.  The $Q$-fluxes, being two T-dualities from $H$-flux, should be invariant under the orbifold group and odd under the orientifold involution, while the $R$-fluxes, being two T-dualities away from metric fluxes, should be invariant under the full orientifold group (a pair of T-dualities should preserve the eigenvalue under world-sheet parity; alternatively, for $Q$-flux we will see this requirement emerge from the base-fiber approach).

It turns out that one can define a natural action of these fluxes on globally defined forms.  So we will again take a $p$-form
\be
A^{(p)}=\frac{1}{p!}A_{i_1\cdots i_p}\eta^{i_1}\w\cdots\w\eta^{i_p},
\ee
with constant coefficients\footnote{We are of course cheating here; when there is no good sense of global geometry, there are no sensible definitions of global forms.  In fact, we hope the reader will view these for now as schematic short-cuts to obtain expressions for the four-dimensional effective theory.  We hope that the role of these constructions becomes clearer in the next section.}, and we will define a map to $(p-1)$-forms,
\be
\lp Q\cdot A\rp_{i_1\cdots i_{p-1}}=\hlf\binom{p-1}{1}Q^{jk}_{[i_1}A_{|jk|i_2\cdots i_{p-1}]},
\ee
and to $(p-3)$-forms,
\be
\lp R\cdot A\rp_{i_1\cdots i_{p-3}}=\frac{1}{6}\binom{p-3}{0}R^{jk\ell}A_{jk\ell i_1\cdots i_{p-3}},
\ee
where we have written the numerical factors in such a way so as to make clear that the forms must be sufficiently high degree ($\ge 2$ for $Q$, $\ge 3$ for $R$) to give a nonzero result.

As before, there will be Bianchi identities restricting the NS-NS fluxes.  There are a number of different approaches to deriving these identities, and we will discuss some of these in Appendix~\ref{BIDerivation}; here we will simply list the results.  We assume that there are no invariant vector fields, so that a tracelessness condition $Q^{ij}_j=0$ is satisfied automatically, analogously to the $\om^j_{ij}=0$ that we demanded previously.  We also assume that there are no zero-forms which are invariant under the orbifold group and odd under the involution, so that $H_{ijk}R^{ijk}=\om^i_{jk}Q^{jk}_i=0$ automatically as well.  With these assumptions, the Bianchi identities read
\bea
H_{m[ij}\om^m_{k\ell]} &=& 0,\non\\
H_{m[ij}Q^{m\ell}_{k]}-\om^m_{[ij}\om^\ell_{k]m} &=& 0,\non\\
H_{ijm}R^{k\ell m}+\om^m_{ij}Q^{k\ell}_m-4\om^{[k}_{m[i}Q^{\ell]m}_{j]} &=& 0,\\
\om^{[j}_{mi}R^{k\ell]m}-Q^{[jk}_mQ^{\ell]m}_i &=& 0,\non\\
Q^{[ij}_mR^{k\ell]m} &=& 0,\non
\eea

It turns out to be very natural to define a sort of covariant differential~\cite{Micu:2007rd}, in analogy to the twisted differential $d_H$ in the case with metric fluxes,
\be
\mathcal{D}=H\w\cdot-\om\cdot+Q\cdot-R\cdot.
\ee
We postpone a full discussion of T-duality until section~\ref{BFApproach}, but the main argument in favor of this formulation is that it appears in the correct T-duality-invariant formulation of the tadpole condition, which will now read as\footnote{Of course, in the absence of a global (or even local) ten-dimensional geometry it is difficult to interpret this expression, which essentially is derived by T-duality, and in particular the interpretation of branes must be subtle~\cite{Lawrence:2006ma,Ellwood:2006my}.}
\be
-\sqrt 2\mathcal{D}F_{RR}+\ls\d_{D6}\rs=2\ls\d_{O6}\rs.
\ee

Furthermore, as discussed more carefully in Appendix~\ref{BIDerivation}, the Bianchi identities above follow from imposing $\mathcal{D}^2=0$ (along with our assumptions).  Finally, the superpotential can also be obtained by duality arguments~\cite{Aldazabal:2006up} (or see the approach of~\cite{Benmachiche:2006df}) and is given by
\be
W=\int_Xe^{J_c}\w F_{RR}+\int_X\Om_c\w\mathcal{D}\lp e^{-J_c}\rp,
\ee
as expected by comparing with (\ref{omegaWQ}).

Now we shall represent our fluxes more succinctly by expansions similar to those we had previously,
\be
Q\cdot\widetilde\om_a=q_{aK}b_K,\qquad Q\cdot\widetilde\m_\al=\widehat q_{\al K}a_K,
\ee
\be
R\cdot\phi=s_Kb_K.
\ee
Though we won't prove it here, we also have
\be
Q\cdot a_K=-\lp d^{-1}\rp^{ab}q_{bK}\om_a,\qquad Q\cdot b_K=\lp{\widehat d}^{-1}\rp^{\al\beta}\widehat q_{\beta K}\m_\al,
\ee
\be
R\cdot a_K=\left|\G\right|s_K\,1,\qquad R\cdot b_K=0.
\ee

Many of the Bianchi identities can be obtained by demanding that $\mathcal{D}^2=0$ on our cohomological basis, namely
\bea
\widehat r_{\al K}p_K=\widehat r_{\al K}s_K=\widehat q_{\al K}p_K=\widehat q_{\al K}s_K &=& 0,\qquad\forall\al,\non\\
\widehat r_{\al K}r_{bK}=\widehat r_{\al K}q_{bK}=\widehat q_{\al K}r_{bK}=\widehat q_{\al K}q_{bK} &=& 0,\qquad\forall\al,b,\non\\
|\G|p_{[K}s_{J]}+\lp d^{-1}\rp^{ab}r_{a[K}q_{|b|J]}=\lp\widehat d^{-1}\rp^{\al\beta}\widehat r_{\al[K}\widehat q_{|\beta|J]} &=& 0,\qquad\forall K,J.\label{BIFromCohomology}
\eea
Unfortunately, as before, there are some Bianchi identities which are not captured by these equalities.

With these definitions, the tadpole condition reads
\be
\label{NonGeometricTadpole}
-\sqrt 2\lp p_Km_0-r_{aK}m_a+q_{aK}e_a-s_Ke_0\rp+N^{(D6)}_K=2N^{(O6)}_K,
\ee
and the superpotential is
\bea
W &=& \frac{1}{|\G|}e_0+d_{ab}t_ae_b+\hlf\k_{abc}t_at_bm_c+\frac{1}{6}m_0\k_{abc}t_at_bt_c\\
&& +2N_K\lp p_K+r_{aK}t_a+\hlf\k_{abc}\lp d^{-1}\rp^{ce}q_{eK}t_at_b+\frac{|\G|}{6}s_K\k_{abc}t_at_bt_c\rp.\non
\eea
The K\"ahler potential $K$ and holomorphic couplings $f_{\al\beta}$ remain unchanged.

The corresponding F-terms are
\bea
\label{NonGeometricFTerm}
D_aW &=& d_{ab}e_b+\k_{abc}m_bt_c+\hlf m_0\k_{abc}t_bt_c+2N_Kr_{aK}+2N_K\k_{abc}\lp d^{-1}\rp^{cd}q_{dK}t_b\non\\
&& +|\G|N_Ks_K\k_{abc}t_bt_c+\frac{3i}{2}\frac{\k_{abc}v_bv_c}{\k_{def}v_dv_ev_f}W,\\
D_KW &=& 2p_K+2r_{aK}t_a+\k_{abc}\lp d^{-1}\rp^{cd}q_{dK}t_at_b+\frac{|\G|}{3}s_K\k_{abc}t_at_bt_c-4e^D\mathcal{F}_KW.\non
\eea

Note that the superpotential only depends on the NS-NS fluxes $p_K$, $r_{aK}$, $q_{aK}$, and $s_K$, and not on the hatted fluxes $\widehat r_{\al K}$ and $\widehat q_{\al K}$.  These appear only in the D-terms, which we turn to next.

\subsubsection{D-terms revisited}
\label{DTermsRevisited}

Recall that in our previous discussion of D-terms, we noted that a gauge transformation of the three-form potential could also, in the presence of certain metric fluxes, shift the axion fields $\xi_K$, because the forms $\m_\al$, on which the three-form is reduced to give the four-dimensional vectors, were no longer closed.  This analysis should still hold in the presence of non-geometric fluxes, but it seems natural to rewrite the extra contribution to the gauge transformation as
\be
\mathcal{D}\lp\lambda^\al\m_\al\rp=-\lambda^\al\widehat r_{\al K}a_K.
\ee
For this gauge transformation, replacing the usual differential $d$ by $\mathcal{D}$ makes no difference; $d\m_\al=\mathcal{D}\m_\al=-\widehat r_{\al K}a_K$.

However, we can also consider gauge transformations of the {\it dual} gauge fields in four dimensions.  These fields are obtained by reducing the five-form potential against a four-form.  Since the orientifold action requires $C_5$ to be odd, we must reduce it against an odd four-form $\widetilde\m_\al$ to get an invariant vector in four-dimensions, $C_5=\widetilde A^\al\w\widetilde\m_\al$.  Then we claim that the dual gauge transformations are generated by gauge transformations of the five-form, and that in principle we can pick up an extra piece,
\be
C_{RR}\longrightarrow C_{RR}+d\widetilde\lambda^\al\w\widetilde\m_\al+\widetilde\lambda^\al\widehat q_{\al K}a_K,
\ee
where $C_{RR}=C_1+C_3+C_5+C_7+C_9$ is a formal sum of R-R potentials.

Thus the fields $\xi_K$ are not invariant under the dual gauge transformation.  In other words, the fields $\Xi_K$ defined in (\ref{XiK}) have magnetic charge $\hlf\widehat q_{\al K}$ under the gauge group $\U(1)_\al$.  These fields can in fact be dyons, that is have both electric and magnetic charges.  It is interesting to ask whether our collection of charged scalars are then mutually local, in the sense of~\cite{Argyres:1995jj} (if they weren't then we would despair of having any Lagrangian description for our effective physics).  The condition that two dyons labelled $K$ and $J$ with these charges be mutually local is simply that
\be
\lp\widehat d^{-1}\rp^{\al\beta}\lp\widehat r_{\al K}\widehat q_{\beta J}-\widehat r_{\al J}\widehat q_{\beta K}\rp=0.
\ee
But this is precisely one of the Bianchi identities of (\ref{BIFromCohomology}), so our fields are guaranteed to be mutually local.

This in turn implies that all charges can be made electric charges by a symplectic transformation $M\in\Sp(2n_V;\Z)$, where $n_V=h^{1,1}_+$ is the number of vectors, and where the symplectic group is defined here by
\be
M\lp\begin{matrix}0 & \widehat d \\ -\widehat d^T & 0\end{matrix}\rp M^T=\lp\begin{matrix}0 & \widehat d \\ -\widehat d^T & 0\end{matrix}\rp,
\ee
and $M$ acts on the $K^{\mathrm{th}}$ charge vector as
\be
M\lp\begin{matrix}-\widehat r_{\al K} \\ \widehat q_{\al K}\end{matrix}\rp=\lp\begin{matrix}-\widehat r_{\al K}' \\ 0\end{matrix}\rp.
\ee
We then have the D-term being
\be
\label{NonGeometricDTerm}
D_\al=-2ie^D\mathcal{F}_K\widehat r'_{\al K}.
\ee

We have already seen that the holomorphic couplings for the electric gauge groups are given by $f^{(\mathrm{electric})}_{\al\beta}=i\lp\widehat\k t\rp_{\al\beta}$, where we have made the obvious definition $\lp\widehat\k t\rp_{\al\beta}=\widehat\k_{\al\beta a}t_a$.  What about for the magnetic gauge groups?  In other words, if the $\widehat r$ fluxes are taken to vanish, but the $\widehat q$ fluxes are nonvanishing then we need to use the dual field strengths in our $\mathcal{N}=1$ description and this will give a different answer for the holomorphic coupling constants.  The new constants can be obtained by rewriting the ten-dimensional action (\ref{MassiveIIAAction}) (with NS-NS fluxes turned off) in terms of the potential $C_5$ rather than its dual $C_3$.  The result is
\be
f^{(\mathrm{magnetic})}_{\al\beta}=-i\lp\widehat\k t\rp^{-1\,\g\d}\widehat d_{\g\al}\widehat d_{\d\beta}.
\ee
This result is also consistent with T-duality.  For the general case, we should take the matrix
\be
f=\lp\begin{matrix}f^{(\mathrm{electric})} & 0 \\ 0 & f^{(\mathrm{magnetic})}\end{matrix}\rp,
\ee
and transform it under our symplectic transformation to get $f'=MfM^T$, and finally take our holomorphic couplings to be $f^{\prime\,(\mathrm{electric})}_{\al\beta}$ (i.e. the top-left block of the matrix $f'$).

By using a combination of $\widehat r$ and $\widehat q$ fluxes, it is apparent that we can get the D-term piece of the potential,
\bea
V_D &=& -2e^{2D}\lp\Re f'\rp^{-1\,\al\beta}\mathcal{F}_K\widehat r'_{\al K}\mathcal{F}_J\widehat r'_{\beta J}\\
&=& -2e^{2D}\ls\lp\Re f^{(\mathrm{electric})}\rp^{-1\,\al\beta}\widehat r_{\al K}\widehat r_{\beta J}+\lp\Re f^{(\mathrm{magnetic})}\rp^{-1\,\al\beta}\widehat q_{\al K}\widehat q_{\beta J}\rs\mathcal{F}_K\mathcal{F}_J,\non
\eea
to have fairly complicated dependence on the K\"ahler moduli.  It would be worthwhile to investigate whether this allows us to achieve some phenomenologically interesting models, with the D-terms either allowing meta-stable deSitter solutions, or possibly even nice inflationary potentials, and we are currently looking at these issues.

\subsubsection{Example}
\label{NonGeometricExample}

Let us briefly see how some of these results work in our example.  Unlike in previous subsections, we won't expend much effort trying to solve the equations, but will content ourselves simply with classifying the fluxes permitted by the orientifold action, and stating the equations that we would like to solve.

The $Q$-fluxes which survive the orientifold projection are
\bea
&Q^{13}_5&=-Q^{14}_5=-Q^{23}_5=-Q^{24}_5,\non\\
&Q^{13}_6&=-Q^{24}_6,\non\\
&&\quad Q^{14}_6=Q^{23}_6=-Q^{13}_5+Q^{13}_6,\non\\
2&Q^{15}_1&=-Q^{16}_1=-2\, Q^{25}_2=Q^{26}_2,\non\\
&Q^{15}_2&=Q^{25}_1,\non\\
&Q^{15}_3&=-Q^{25}_4,\non\\
&Q^{16}_3&=Q^{16}_4=Q^{26}_3=-Q^{26}_4,\\
&&\quad Q^{15}_4=Q^{25}_3=-Q^{15}_3-Q^{16}_3,\non\\
&Q^{35}_1&=-Q^{45}_2,\non\\
&Q^{36}_1&=Q^{36}_2=Q^{46}_1=-Q^{46}_2,\non\\
&&\quad Q^{35}_2=Q^{45}_1=-Q^{35}_1-Q^{36}_1,\non\\
&Q^{35}_3&=-Q^{45}_4,\non\\
2&Q^{35}_4&=-Q^{36}_4=2\, Q^{45}_3=-Q^{46}_3,\non
\eea
where we take the ten fluxes in the left-hand column as independent.  Similarly there are two independent $R$-fluxes,
\bea
&R^{135}&=-R^{245},\non\\
&R^{136}&=-R^{146}=-R^{236}=-R^{246},\\
&&\quad R^{145}=R^{235}=R^{135}+R^{136}.\non
\eea

In more succinct terms,
\be
q=\lp\begin{matrix}-Q^{35}_1 & -Q^{35}_1-Q^{36}_1 \\ Q^{15}_3 & Q^{15}_3+Q^{16}_3 \\ -Q^{13}_5+Q^{13}_6 & Q^{13}_6 \\ Q^{15}_1-Q^{15}_2-Q^{35}_3-Q^{35}_4 & -Q^{15}_1-Q^{15}_2-Q^{35}_3+Q^{35}_4\end{matrix}\rp,
\ee
\be
\widehat q=\lp\begin{matrix}-Q^{15}_1+Q^{15}_2-Q^{35}_3-Q^{35}_4 & \  & -Q^{15}_1-Q^{15}_2+Q^{35}_3-Q^{35}_4\end{matrix}\rp,
\ee
\be
s=\lp\begin{matrix}R^{135}+R^{136} & \  & R^{135}\end{matrix}\rp.
\ee

The Bianchi identities are unfortunately quite complicated and unenlightening.  In addition to the identities from (\ref{BIFromCohomology}), we have the following extra conditions:
\be
-8r_{31}s_1+8q_{11}q_{21}-\lp q_{41}\rp^2-\lp\widehat q_1\rp^2=-8r_{32}s_2+8q_{12}q_{22}-\lp q_{42}\rp^2-\lp\widehat q_2\rp^2,\non
\ee
\be
\k_{3ab}\lp d^{-1}\rp^{cb}r_{c(K}s_{J)}=q_{3(K}q_{|a|J)},\qquad\mathrm{for}\ a=1,2,4;\ \forall K,J,\non
\ee
\be
s_1\widehat r_2+s_2\widehat r_1=\widehat q_1q_{32}+\widehat q_2q_{31},\non
\ee
\be
4q_{11}r_{11}+4q_{21}r_{21}-q_{41}r_{41}-\widehat q_1\widehat r_1-8q_{31}r_{31}=4q_{12}r_{12}+4q_{22}r_{22}-q_{42}r_{42}-\widehat q_2\widehat r_2-8q_{32}r_{32},\non
\ee
\be
\label{ExampleBI}
r_{a(K}\widehat q_{J)}=\k_{3ab}\lp d^{-1}\rp^{cb}q_{c(K}\widehat r_{J)},\qquad\mathrm{for}\ a=1,2,4;\ \forall K\ne J,
\ee
\be
\k_{3ac}\lp d^{-1}\rp^{dc}q_{d(K}r_{|b|J)}=\k_{3bc}\lp d^{-1}\rp^{dc}q_{d(K}r_{|a|J)},\qquad\mathrm{for}\ a,b\in\{1,2,4\},\ \forall K,J,\non
\ee
\be
q_{3(K}r_{|3|J)}+p_{(K}s_{J)}=0,\non
\ee
\be
-8p_1q_{31}+8r_{11}r_{21}-\lp r_{41}\rp^2-\lp\widehat r_1\rp^2=-8p_2q_{32}+8r_{12}r_{22}-\lp r_{42}\rp^2-\lp\widehat r_2\rp^2,\non
\ee
\be
\k_{3ab}\lp d^{-1}\rp^{cb}q_{c(K}p_{J)}=r_{3(K}r_{|a|J)},\qquad\mathrm{for}\ a=1,2,4;\ \forall K,J,\non
\ee
\be
p_1\widehat q_2+p_2\widehat q_1=\widehat r_1r_{32}+\widehat r_2r_{31}.\non
\ee

The tadpole conditions are just as listed in (\ref{NonGeometricTadpole}), the D-term equations require (\ref{NonGeometricDTerm}) to vanish, and the F-term equations are as given in (\ref{NonGeometricFTerm}).

\subsection{Summary and Puzzles}
\label{Puzzles}

We have laid out an approach to studying a class of four-dimensional $\mathcal{N}=1$ effective theories.  Starting from toroidal orientifolds of IIA string theory with NS-NS $H$-flux turned on, we followed in the footsteps of many authors before us and argued for a more general class of NS-NS fluxes.  The arguments proceed roughly by showing at each step that a T-duality induces the possibility of a new type of flux, and then we generalize to a framework capable of accomodating these new fluxes as well as the old ones (and thus allowing configurations that are not simply T-dual to previous ones).  In this way we introduced metric fluxes $\om^i_{jk}$, then non-geometric fluxes $Q^{ij}_k$, and finally $R^{ijk}$.

However, these arguments were really made at the level of the effective field theory.  In terms of ten-dimensional constructions, there would seem to be some obstacles to this program.  For instance, beginning with $H$-flux on a torus, say $h\,dx\w dy\w dz$, to perform a ten-dimensional T-duality, one first picks a trivialization of the $B$-field such as $B=hx\,dy\w dz$.  Then the Buscher rules~\cite{Buscher:1987sk} allow one to T-dualize along either the $y$ or $z$ directions, resulting in metric flux $\om^y_{xz}$ or $\om^z_{xy}$, or T-dualize in $y$ and $z$, resulting in $Q^{yz}_x$, but it is not obvious how to perform the third T-duality here to get $R^{xyz}$; our trivialization broke the third isometry, and the Buscher rules no longer apply.  Indeed, there are general arguments that any ten-dimensional origin for $R$-flux cannot even have a local description~\cite{Shelton:2006fd,Bouwknegt:2004ap}.  So it is very much of interest to ask which configurations can be constructed from ten dimensions.

We have also tried to formulate everything in a language that moves away from the toroidal context.  So, instead of phrasing everything in terms of flux components, $H_{ijk}$, $\om^i_{jk}$, $Q^{ij}_k$ and $R^{ijk}$, we rewrite our formulae (thereby serendipitously simplifying the $\mathcal{N}=1$ expressions at the same time) in terms of matrices $p_K$, $r_{aK}$, $\widehat r_{\al K}$, $q_{aK}$, $q_{\al K}$ and $s_K$ which referred only to the (untwisted) cohomology of the orientifold.  Our hope is that this language will also allow the study of general NS-NS fluxes on arbitrary type IIA Calabi-Yau orientifold constructions, resulting in a greatly enriched tool-kit for model building.  The major flaw right now in this plan is the Bianchi identities, which we were unable, in general, to recast in terms of the cohomological structure alone.  Our hope, however, is that this difficulty can be overcome by studying explicit examples.

The most obvious extension in this direction would be to address another fairly prominent gap in our analysis, namely the incorporation of the twisted sectors.  We have ignored twisted sector fluxes and moduli throughout our analysis, since we are more interested, in the present work, in elucidating the general structures that one encounters.  In other contexts~\cite{DeWolfe:2005uu}, it has been shown that, at least in specific models, it should be possible to stabilize the twisted sector moduli in such a way as to maintain a separation of scales with the bulk physics, but still trust the analysis.  We hope that such considerations will still hold in many of the models discussed here.  It would be very interesting to incorporate the twisted-sector cohomology into our general flux analysis (indeed if our approach is valid beyond toroidal examples, then it should be able to treat all of the cohomology democratically), possibly along the lines of~\cite{Cvetic:2007ju}.  In section~\ref{AdvantagesAndPuzzles} we will mention some ideas in this direction.

Another key point to emphasize here is the quantization of general NS-NS fluxes.  For $H$-fluxes alone, the situation is well understood; the $H$-flux should be understood as an element of $H^3(X;\Z)$, or in our terms, the $p_K$ should be integers\footnote{Actually, related to our willful ignorance of the twisted sectors, we have glossed over the fact that in our example, $p_K$ should in fact be even integers~\cite{Blumenhagen:2002gw}; our $b_K$ alone are not elements of the integral cohomology, but rather we must take either $nb_1+mb_2$ with $n+m$ even, or we may take $b_K+(\mathrm{twisted})$.  It would be interesting to provide a more complete analysis.}.  In situations related to these by T-duality the answers are just as straightforward; all of the fluxes $p_K$, $r_{aK}$, etc. must be integers.  It is natural to assume that this is generally the correct condition, especially when we are describing our fluxes in terms of integral cohomology.  However, as we shall see in the next section, this naive quantization is not generally correct.  There will be examples we can construct (which are not simply T-dual to $H$-flux) where the quantization condition is much more complicated (though still simple from the point of view of our constructions).  This still leaves the question of how fluxes are quantized in those models that we will not succeed in constructing from a ten-dimensional point of view.  In that case we do not know what the correct quantization conditions should be.  It is possible that those models simply have no legitimate ten-dimensional origin.  If they do, we see no route to determining the correct quantization conditions without actual constructions.

In section~\ref{omegaExample} we presented one example of a model where all moduli were stabilized at a supersymmetric AdS vacuum and the tadpole condition was saturated without the need for extra D-branes.  Unfortunately, this example used only our naive quantization conditions.  Using the correct quantization on NS-NS fluxes which we will derive below we will find that it is no longer possible to stabilize all moduli while also satisfying both the F-term equations and the tadpole, the latter because the flux contributions in this case appear to be non-integral!  We suspect that this problem with the tadpole is simply an artifact of our not understanding how the generalized NS-NS fluxes affect the correct quantization of R-R fluxes.  It would be extremely gratifying to have a better grasp of these issues so as to be able to construct fully realized stable $\mathcal{N}=1$ vacua\footnote{Of course in section~\ref{omegaExample}, since we have only turned on $H$-flux and metric flux, we do still have a global geometric description, and there should be nothing exotic about the quantization of R-R fluxes.  Our suspicion, however is that we have run into trouble by trying to use the language of the twisted torus, i.e. in using fluxes defined by forms inherited from $T^6$.  For the types of metric flux used here (and similar examples in the literature), the resulting space is quite different from the original $T^6$, and so the quantization conditions in our chosen basis will seem quite non-standard.\label{RRQuantizationFootnote}}.

Finally, let us turn to the issue of the regime of validity of this effective field theory.  As in~\cite{DeWolfe:2005uu}, we are able to find models (by taking some of our R-R fluxes to be parametrically large, for instance $\widehat e_a$ in our solutions with $H$-flux only) in which the string coupling is small, and in which the compact directions are large enough to trust supergravity, but still much smaller than the AdS radius (which also characterizes the masses of the stabilized moduli), so that the solution would seem to be effectively four-dimensional\footnote{Note that these conditions are not preserved by T-duality.}.  However, just as in that situation, our models generally suffer from the concerns expressed by Banks and van den Broek~\cite{Banks:2006hg}.  Namely, due to the presence of the orientifold singularity, there are regions of our compact manifold in which the string coupling diverges (but see also~\cite{Acharya:2006ne}) and we should turn to eleven-dimensional supergravity instead.  In this picture, the large flux integers translate into a large stack of M2-branes at the orientifold locus, and so the larger the flux integers, the more backreaction one has to deal with (and is ignoring in the effective description).  We have not repeated this analysis in detail in our models, partly because the ten-dimensional (or eleven-dimensional) physics becomes more obscure for us, but the issue undoubtedly persists.  We hope however, that our richer structure of fluxes might provide more corners in which to hide.

\section{Base-Fiber Approach}
\label{BFApproach}

In this section we will attempt to put a subset of our class of models on firmer ground by presenting ten-dimensional constructions.  These constructions are very much in the spirit of~\cite{Dabholkar:2002sy} (see also~\cite{Hellerman:2002ax,Flournoy:2004vn,Hull:2004in,Dabholkar:2005ve,Hull:2006qs,Hull:2006va}) and are built by allowing a torus fiber to vary over a torus base, but in a way that still admits a generalized Scherk-Schwarz reduction.  The NS-NS fluxes will be represented by the global twists in the fibers as one transports them around non-contractible cycles in the base.  We will find that the Bianchi identities come out naturally, that dualities are implemented very easily, and that the correct quantization conditions are both obvious in this context, and also much more subtle than one would have guessed.

\subsection{The T-duality group $O(6,6;\Z)$}
\label{TheTDualityGroup}

The T-duality group of type II superstring theory compactified on a $d$-dimensional torus $T^d$ is denoted $\O(d,d;\Z)$ and is defined as follows
\be
\label{OddZDef}
\O(d,d;\Z)=\left\{M\in\operatorname{Mat}_{2d\times 2d}(\Z)|MLM^T=L\right\},
\ee
where
\be
L=\begin{pmatrix}0 & {\bf 1}_d \\ {\bf 1}_d & 0\end{pmatrix}.
\ee
We will in fact focus primarily on elements with determinant one, which correspond to dualities from IIA to itself (or IIB to itself); elements with determinant minus one interchange solutions of IIA and IIB.

To understand the action of this group on the NS-NS sector, it is convenient to combine the torus metric and $B$-field into a single $d\times d$ matrix $E=G+B$.  We assume implicitly here that our coordinate basis is chosen such that each coordinate is periodic with unit period.  Let us take an $\O(d,d;\Z)$ matrix $M$ and write it in terms of $d\times d$ blocks,
\be
M=\lp\begin{matrix}a & b \\ c & d\end{matrix}\rp.
\ee
Then the action of $M$ on the NS-NS sector is
\be\label{OddNSNS}
E\mapsto E'=\lp aE+b\rp\lp cE+d\rp^{-1},\qquad e^\phi\mapsto e^{\phi'}=e^\phi\lp\frac{\det G'}{\det G}\rp^{1/4}.
\ee
There is a useful alternative phrasing of this transformation.  From $G$ and $B$ we can define a symmetric $2d\times 2d$ matrix
\be
H=\lp\begin{matrix}G-BG^{-1}B & BG^{-1} \\ -G^{-1}B & G^{-1}\end{matrix}\rp.
\ee
Then an element $M\in\O(d,d;\Z)$ simply acts by
\be
\label{OddNSNSH}
H\mapsto H'=M^THM.
\ee

From this we can identify certain important elements of $\O(d,d;\Z)$.  For instance, it includes changes of basis for the lattice which defines $T^d$.  These basis changes lie in the subgroup $\GL(d;\Z)\subset\O(d,d;\Z)$ of matrices with the form
\be\label{GLdZinOdd}
\hat g=\begin{pmatrix}\lp g^T\rp^{-1} & 0 \\ 0 & g\end{pmatrix},\qquad g\in\GL(d;\Z).
\ee
Similarly, we also have constant integral shifts in the periods of the $B$-field given by matrices
\be
\begin{pmatrix}{\bf 1}_d & b \\ 0 & {\bf 1}_d\end{pmatrix},\qquad b^T=-b.
\ee

Finally there is one more type of element which will be of interest to us, corresponding simply to T-dualizing a sub-torus $T^k$ of $T^d$, for example that corresponding to the first $k$ coordinates.  Then the relevant $M$ is
\be
\label{Mk}
M_k=\begin{pmatrix}0 & 0 & {\bf 1}_k & 0 \\ 0 & {\bf 1}_{d-k} & 0 & 0 \\ {\bf 1}_k & 0 & 0 & 0 \\ 0 & 0 & 0 & {\bf 1}_{d-k}\end{pmatrix}.
\ee
From this and the transformation rules (\ref{OddNSNS}) above, one can compute the usual Buscher rules.

It will also be useful to know how elements of $\O(d,d)$ act on the R-R fluxes and potentials, which can be thought of as sections of the spin bundle $\operatorname{Spin}(d,d)$.  The action in this case cannot be expressed as simply as in the cases above~\cite{Hassan:1999mm}, but it is not hard to write down for certain simple cases which can then be used to generate all of $\O(d,d)$ \cite{Gualtieri:2003dx}.  In particular, we have three cases.

If $B_{ij}$ is an antisymmetric $d\times d$ matrix, so that $B=\hlf B_{ij}dx^i\w dx^j$ is a two-form on the torus, then the element
\be
g=\lp\begin{matrix}{\bf 1}_d & B \\ 0 & {\bf 1}_d\end{matrix}\rp
\ee
acts as
\be
g\cdot F_{RR}=\exp\lp B\rp\w F_{RR}=F_{RR}+B\w F_{RR}+\hlf B\w B\w F_{RR}+\cdots,
\ee
where $F_{RR}=\sum_aF_a$ is the sum of R-R-fluxes of various degrees.

Similarly, if $\beta^{ij}$ is antisymmetric, $\beta=\hlf\beta^{ij}\frac{\p}{\p x^i}\frac{\p}{\p x^j}$ an antisymmetric bivector, then
\be
\lp\begin{matrix}{\bf 1}_d & 0 \\ \beta & {\bf 1}_d\end{matrix}\rp\cdot F_{RR}=\exp\lp\iota_\beta\rp\cdot F_{RR},
\ee
where $\iota_\beta=\hlf\beta^{ij}\iota_{\p_i}\iota_{\p_j}$ and $\iota_v$ acts by contracting a form with a vector $v$.

Finally, if $g\in\GL(d)$, then
\be
\lp\begin{matrix}\lp g^T\rp^{-1} & 0 \\ 0 & g\end{matrix}\rp\cdot F_{RR}=\left|\det g\right|^{1/2}\lp g^{-1}\rp^\ast F_{RR},
\ee
where $(g^{-1})^\ast F_{RR}$ denotes the pullback of $F_{RR}$ by the map $g^{-1}$.

We are primarily interested in studying toroidal orientifolds, so it is important to understand how to discuss the orientifold group action in this language.  For elements of the orbifold group, this is fairly clear; for any lattice preserving diffeomorphism $g\in\GL(d;\Z)$ which acts on our torus, such as a rotation, we simply need to construct the corresponding element $\hat g\in\O(d,d;\Z)$ as in (\ref{GLdZinOdd}) above.  To describe the full orientifold action, we also need to know how the world-sheet parity operator $\Om$ acts.  It turns out that $\Om$ can also be expressed as a $2d\times 2d$ matrix,
\be
\label{Omega}
\Om=\begin{pmatrix}-{\bf 1}_d & 0 \\ 0 & {\bf 1}_d\end{pmatrix},
\ee
with the understanding that this operator also acts on the remaining $10-d$ coordinates (so that, e.g. it does not exchange IIA and IIB, even if $d$ is odd).  This is not an element of $\O(d,d;\Z)$, since it satisfies that $\Om L\Om^T=-L$ rather than (\ref{OddZDef}), but it can be thought of as an element of $\operatorname{Spin}(d,d;\Z)$ and we can understand its action on NS-NS moduli simply by following (\ref{OddNSNSH}), i.e. $\Om\cdot G=G$, $\Om\cdot B=-B$.  We can also work out the action of $\Om$ on R-R fields by following~\cite{Gualtieri:2003dx}, but we in fact know the answer; $C_3$ and $C_7$, as well as $F_0$ and $F_4$ should be even, while $C_1$ and $C_5$, as well as $F_2$ and $F_6$, should be odd.  In this way, we see that the entire orientifold group can be understood as a finite subgroup of $\operatorname{Spin}(d,d;\Z)$ (for instance in our example this subgroup would be generated in this notation by $\hat\Theta$ and $\Omega\hat\s$).

In this notation, the untwisted moduli are simply those which are fixed by the orientifold subgroup $\widehat\G\subset\operatorname{Spin}(d,d;\Z)$.  Note that so far we have not discussed any NS-NS fluxes.  R-R fluxes can be accomodated, and both the R-R fluxes and R-R potentials are understood to transform according to the rules described above.

Now given any element $h\in\SO(d,d;\Z)\subset\operatorname{Spin}(d,d;\Z)$,\footnote{Though we won't use them here, we can certainly in general consider dualities $h$ which lie in $\O(d,d;\Z)\subset\operatorname{Pin}(d,d;\Z)$ and take us from IIA to IIB and vice versa.  It will still be true that $h\widehat\G h^{-1}\subset\operatorname{Spin}(d,d;\Z)$.} we can relate a given orientifold with subgroup $\widehat\G$ and moduli given by $H$, etc., to a dual orientifold with subgroup $h\widehat\G h^{-1}$ and moduli given by $h^THh$, etc.  Note that in general the elements in $h\widehat\G h^{-1}$ need not be block diagonal; the dual orientifold group can be an asymmetric orientifold.

We would like to actually use dualities as a solution generating technique.  In this case we focus on elements which do not modify the orientifold group, i.e. the set of $h\in\SO(d,d;\Z)$ that satisfy $h\widehat\G=\widehat\G h$.  We can consider such $h$ as simply a map on the moduli and fluxes.

\subsubsection{Example}
\label{TDualityGroupExample}

Let us see how this works in our example.  There our orientifold is generated by $\hat\Theta$ and $\Om\hat\s$, where
\be
\label{ThetaAndSigma}
\Theta=\begin{pmatrix}\begin{smallmatrix}0 & -1 \\ 1 & 0\end{smallmatrix} & 0 & 0 \\ 0 & \begin{smallmatrix}0 & -1 \\ 1 & 0\end{smallmatrix} & 0 \\ 0 & 0 & -{\bf 1}_2\end{pmatrix},\qquad\s=\begin{pmatrix}\begin{smallmatrix}1 & 0 \\ 0 & -1\end{smallmatrix} & 0 & 0 \\ 0 & \begin{smallmatrix}0 & 1 \\ 1 & 0\end{smallmatrix} & 0 \\ 0 & 0 & \begin{smallmatrix}1 & 1 \\ 0 & -1\end{smallmatrix}\end{pmatrix},
\ee
are both elements of $\GL(6;\Z)$.

We would like to identify how to perform a T-duality on (for example) the third two-torus.  Unfortunately, by just using (\ref{Mk}), one finds that $\hat\Theta$ is invariant, but $\Om\hat\s$ is not.  However, one can repair this by combining the standard T-duality with a further rotation $(x_3,y_3)\mapsto(y_3,-x_3)$, defining instead the element
\be\label{MT3}
M_{T(3)}=\begin{pmatrix}{\bf 1}_4 & 0 & 0 & 0 \\ 0 & 0 & 0 & \begin{smallmatrix}0 & 1 \\ -1 & 0\end{smallmatrix} \\ 0 & 0 & {\bf 1}_4 & 0 \\ 0 & \begin{smallmatrix}0 & 1 \\ -1 & 0\end{smallmatrix} & 0 & 0\end{pmatrix}.
\ee
This version of T-duality does indeed preserve the full orientifold group.  

On the NS-NS moduli one can check that it acts by sending $t_3\mapsto-1/t_3$, $e^\phi\mapsto e^\phi/|t_3|$, and all other moduli remain fixed.  To get the action on the R-R fields, it is useful to decompose $M_{T(3)}$ as
\be
M_{T(3)}=\lp\begin{matrix}{\bf 1}_4 & 0 & 0 & 0 \\ 0 & {\bf 1}_2 & 0 & 0 \\ 0 & 0 & {\bf 1}_4 & 0 \\ 0 & \begin{smallmatrix}0 & 1 \\ -1 & 0\end{smallmatrix} & 0 & {\bf 1}_2\end{matrix}\rp\lp\begin{matrix}{\bf 1}_4 & 0 & 0 & 0 \\ 0 & {\bf 1}_2 & 0 & \begin{smallmatrix}0 & 1 \\ -1 & 0\end{smallmatrix} \\ 0 & 0 & {\bf 1}_4 & 0 \\ 0 & 0 & 0 & {\bf 1}_2\end{matrix}\rp\lp\begin{matrix}{\bf 1}_4 & 0 & 0 & 0 \\ 0 & {\bf 1}_2 & 0 & 0 \\ 0 & 0 & {\bf 1}_4 & 0 \\ 0 & \begin{smallmatrix}0 & 1 \\ -1 & 0\end{smallmatrix} & 0 & {\bf 1}_2\end{matrix}\rp.
\ee
Now if one writes, for example,
\be
F_{RR}=F^\perp+dx_3\w F^x+dy_3\w F^y+dx_3\w dy_3\w F_\parallel,
\ee
one finds the T-duality relation,
\be
M_{T(3)}\cdot F_{RR}=-F^\parallel+dx_3\w F^x+dy_3\w F^y+dx_3\w dy_3\w F^\perp.
\ee
As a consequence, we find that the $\xi_K$ are invariant, while the R-R fluxes map according to
\be
m_0'=-m_3,\qquad m_1'=-e_2,\qquad m_2'=-e_1,\qquad m_3'=m_0,\qquad m_4'=-e_4\non
\ee
\be
e_1'=m_2,\qquad e_2'=m_1,\qquad e_3'=-e_0,\qquad e_4'=m_4,\qquad e_0'=e_3,
\ee
where primed quantities represent the fluxes in the T-dual solution and unprimed ones are from the original solution.  We will discuss duality of NS-NS fluxes after introducing them in the next subsection.

We can now similarly introduce T-dualities $M_{T(1)}$ and $M_{T(2)}$, corresponding to dualizing either the first or second two-torus, by simply permuting the two-by-two blocks of $M_{T(3)}$.  Under $M_{T(1)}$ we have $t_1\mapsto-1/t_1$, $t_2\mapsto t_2-2t_4^2/t_1$, $t_4\mapsto-t_4/t_1$, and $e^\phi\mapsto e^\phi/|t_1|$, while 
\be
m_0'=-m_1,\qquad m_1'=m_0,\qquad m_2'=-e_3,\qquad m_3'=-e_2,\qquad m_4'=m_4,\non
\ee
\be
e_1'=-e_0,\qquad e_2'=m_3,\qquad e_3'=m_2,\qquad e_4'=e_4,\qquad e_0'=e_1,
\ee
with all other moduli left invariant.  The action of $M_{T(2)}$ can be obtained from $M_{T(1)}$ by interchanging one and two throughout.

\subsection{NS-NS Fluxes}
\label{NSNSFluxes}

In order to get a feeling for how we would like to encode general NS-NS fluxes, let us start with the example of $T^6/\Z_4$ with only $H$-flux, from section~\ref{HExample}.

\subsubsection{Example}
\label{NSNSFluxExample}

Here we have $H_3=p_1b_1+p_2b_2$.  In order to represent this flux, let us first pick a trivialization that depends only on the coordinates $x_1$ and $y_1$ (these coordinates will then be our {\it base}).  Our $B$-field is thus
\begin{align}
B & =p_1\ls-\lp x_1-y_1\rp dx_2\w dx_3+y_1dx_2\w dy_3+\lp x_1+y_1\rp dy_2\w dx_3+x_1dy_2\w dy_3\rs\non\\
\label{BTriv}
& +p_2\ls\lp x_1-y_1\rp dx_2\w dx_3+x_1dx_2\w dy_3-\lp x_1+y_1\rp dy_2\w dx_3-y_1dy_2\w dy_3\rs.
\end{align}

If we let $E_0$ be the combination of the metric and $B$-field at the point $x_1=y_1=0$ (including values of the moduli $t_a$), then we can write $E(x_1,y_1)=g(x_1,y_1)\cdot E_0$, where $g(x_1,y_1)$ is a map of the base $T^2$ into $\O(4,4)\subset\O(6,6)$ given explicitly by
\be
g(x_1,y_1)=\lp\begin{matrix}{\bf 1}_4 & \begin{smallmatrix}0 & 0 & (p_2-p_1)(x-y) & p_1y+p_2x \\ 0 & 0 & (p_1-p_2)(x+y) & p_1x-p_2y \\ (p_1-p_2)(x-y) & (p_2-p_1)(x+y) & 0 & 0 \\ -p_1y-p_2x & -p_1x+p_2y & 0 & 0\end{smallmatrix} \\ 0 & {\bf 1}_4\end{matrix}\rp=\exp\ls xM_x+yM_y\rs,
\ee
where we have suppressed the subscript $1$ on $x$ and $y$, and where in the final step we have defined
\be
M_x=\lp\begin{matrix}0 & \begin{smallmatrix}0 & 0 & p_2-p_1 & p_2 \\ 0 & 0 & p_1-p_2 & p_1 \\ p_1-p_2 & p_2-p_1 & 0 & 0 \\ -p_2 & -p_1 & 0 & 0\end{smallmatrix} \\ 0 & 0\end{matrix}\rp,\qquad M_y=\lp\begin{matrix}0 & \begin{smallmatrix}0 & 0 & p_1-p_2 & p_1 \\ 0 & 0 & p_1-p_2 & -p_2 \\ p_2-p_1 & p_2-p_1 & 0 & 0 \\ -p_1 & p_2 & 0 & 0\end{smallmatrix} \\ 0 & 0\end{matrix}\rp,
\ee
which are mutually commuting constant elements of the Lie algebra $\mathfrak{so}(4,4)$.  Note that the map $g$ is not single-valued, but that upon going around a closed cycle in the base the transformation needs to be a symmetry, i.e. we must have
\be
g(n,m)\in\O(4,4;\Z), \forall n,m\in\Z\qquad\Leftrightarrow\qquad\exp\lp M_x\rp\in\O(4,4;\Z),\ \exp\lp M_y\rp\in\O(4,4;\Z).
\ee
This is indeed satisfied by these matrices for integer values of $p_I$ (both satisfy $M^2=0$ and hence $\exp(M)=1+M$).

We see that $g(x_1,y_1)$, or equivalently $M_x$ and $M_y$, encodes our $H$-fluxes.  What about metric fluxes?  We would like to see how these fluxes map when we T-dualize on the third two-torus.  Since we know how the metric and $B$-field transform, we have
\be
M_{T(3)}\cdot E(x,y)=\lp M_{T(3)}g(x,y)M_{T(3)}^{-1}\rp\cdot\lp M_{T(3)}\cdot E_0\rp.
\ee
So we see that we should replace our twist $g(x_1,y_1)$ by a new twist in $\O(4,4)\subset\O(6,6)$,
\be
g'(x,y)=M_{T(3)}gM_{T(3)}^{-1}=\lp\begin{matrix}{\bf 1}_2 & \begin{smallmatrix}p_1y+p_2x & (p_1-p_2)(x-y) \\ p_1x-p_2y & (p_2-p_1)(x+y)\end{smallmatrix} & 0 & 0 \\ 0 & {\bf 1}_2 & 0 & 0 \\ 0 & 0 & {\bf 1}_2 & 0 \\ 0 & 0 & \begin{smallmatrix}-p_1y-p_2x & -p_1x+p_2y \\ (p_2-p_1)(x-y) & (p_1-p_2)(x+y)\end{smallmatrix} & {\bf 1}_2\end{matrix}\rp,
\ee
or equivalently,
\be
M_x'=\lp\begin{matrix}0 & \begin{smallmatrix}p_2 & p_1-p_2 \\ p_1 & p_2-p_1\end{smallmatrix} & 0 & 0 \\ 0 & 0 & 0 & 0 \\ 0 & 0 & 0 & 0 \\ 0 & 0 & \begin{smallmatrix}-p_2 & -p_1 \\ p_2-p_1 & p_1-p_2\end{smallmatrix} & 0\end{matrix}\rp,\qquad M_y'=\lp\begin{matrix}0 & \begin{smallmatrix}p_1 & p_2-p_1 \\ -p_2 & p_2-p_1\end{smallmatrix} & 0 & 0 \\ 0 & 0 & 0 & 0 \\ 0 & 0 & 0 & 0 \\ 0 & 0 & \begin{smallmatrix}-p_1 & p_2 \\ p_1-p_2 & p_1-p_2\end{smallmatrix} & 0\end{matrix}\rp.
\ee
Again these are two commuting elements of $\mathfrak{so}(4,4)$ which exponentiate to elements of $\O(4,4;\Z)$, though this time rather than shifting the $B$-field, they act as diffeomorphisms of the fibered $T^4$.  Note that if we write
\be
g'(x,y)=\lp\begin{matrix}\lp h^T\rp^{-1} & 0 \\ 0 & h\end{matrix}\rp,\qquad h\in\SL(4),
\ee
then we have
\be
\eta^i=\lp h^{-1}\rp^i_jdx^j=\lp h^{-1}\rp^\ast dx^i.
\ee
These are the proper, globally-defined one-forms, since as we traverse the base, we are forced to transport our fiber one-forms by the map $g'$.

In the case just described, we can then compute the metric flux components, namely
\be
\om^5_{13}=-\om^5_{24}=-p_2,\qquad\om^5_{14}=\om^5_{23}=-p_1,\qquad\om^6_{13}=-\om^6_{14}=-\om^6_{23}=-\om^6_{24}=p_2-p_1,
\ee
or in terms of an $r$-matrix (compare with (\ref{rMatrixExample}) and (\ref{widehatrMatrixExample}))
\be
r=\lp\begin{matrix}0 & 0 \\ 0 & 0 \\ -p_1 & -p_2 \\ 0 & 0\end{matrix}\rp,\qquad\widehat r=0.
\ee
Thus we see that this particular T-duality has simply sent $p_K\mapsto-r_{3K}$ (also there is no $H$-flux in this new solution described by $g'$).

By combining this map with the map of moduli and R-R fluxes under T-duality from the previous subsection, one can verify that the solution of section~\ref{HExample} and that of (\ref{NoHCaseaSolution}) are precisely T-dual to each other.

Let us now perform one more T-duality using $M_{T(2)}$.  This sends us to
\be
M_x''=\lp\begin{matrix}0 & 0 \\ \begin{smallmatrix}0 & 0 & p_1 & p_2-p_1 \\ 0 & 0 & -p_2 & p_2-p_1 \\ -p_1 & p_2 & 0 & 0 \\ p_1-p_2 & p_1-p_2 & 0 & 0\end{smallmatrix} & 0\end{matrix}\rp,\qquad M_y''=\lp\begin{matrix}0 & 0 \\ \begin{smallmatrix}0 & 0 & -p_2 & p_2-p_1 \\ 0 & 0 & -p_1 & p_1-p_2 \\ p_2 & p_1 & 0 & 0 \\ p_1-p_2 & p_2-p_1 & 0 & 0\end{smallmatrix} & 0\end{matrix}\rp.
\ee
This will correspond to nongeometric $Q$-flux.  We will argue below in the general case how one should convert these to particular components of $Q$-flux; for now we merely state the results.
\be
Q^{35}_1=-Q^{45}_2=-p_1,\qquad Q^{36}_1=Q^{36}_2=Q^{46}_1=-Q^{46}_2=p_1-p_2,\qquad Q^{35}_2=Q^{45}_1=p_2,
\ee
or
\be
q=\lp\begin{matrix}p_1 & p_2 \\ 0 & 0 \\ 0 & 0 \\ 0 & 0\end{matrix}\rp,\qquad\widehat q=0.
\ee
By applying the $M_{T(2)}$ map to the moduli and fluxes of (\ref{NoHCaseaSolution}) we can generate a new solution with only $Q$-flux (no $H$-flux or metric flux).  It is straightforward to check that the F-term equations are satisfied.

\subsubsection{General situation}
\label{GeneralSituation}

Let us attempt to generalize this situation.  Let $\widehat\G$ be the subgroup of $\operatorname{Spin}(6,6;\Z)$ which generates our orientifold group, and suppose we have a splitting of our $T^6$ into a base of dimension $n$ and a fiber of dimension $6-n$ such that $\widehat\G$ acts block diagonally (i.e. such that both the base and fiber form real representations of the orientifold group, which is $D_4$ in our example).  We will also assume that $\widehat\G$ acts symmetrically on the base, with each element giving rise to a $\GL(n;\Z)$ action.  Then for every element $h\in\widehat\G$ we can decompose
\be
h=h_b\oplus h_f\in\GL(n;\Z)\times\operatorname{Spin}(6-n,6-n;\Z)\subset\operatorname{Spin}(6,6;\Z).
\ee
We would like to classify the elements $g(\vec x_b)\in\SO(6-n,6-n)\subset\SO(6,6)$, depending on the base coordinates $\vec x_b$, by which we can twist our fibers.  Such twists will have to satisfy a number of conditions.

First of all, they need to be invariant under the orientifold group action, i.e. we require
\be
h_fg(\vec x_b)h_f^{-1}=g(h_b\cdot\vec x_b),\qquad\forall h\in\widehat\G.
\ee

Secondly, we require path independence, in the sense that moving in different directions in the base should commute, i.e.
\be
\ls\p_ig(\vec x_b),\p_jg(\vec x_b)\rs,\qquad\forall i,j\in\{1,\ldots,n\}.
\ee

And finally, moving around a closed path must correspond to an element of the duality group, i.e.
\be
g(\vec x_b+\vec\lambda)g(\vec x_b)^{-1}\in\SO(6,6;\Z),\qquad\forall\vec\lambda\in\Z^6,\ \vec x_b\in\R^n\subset\R^6.
\ee

A very natural simplifying ansatz for the form of $g(\vec x_b)$ is to take
\be
g(\vec x_b)=\exp\ls\vec x_b\cdot\vec M\rs,
\ee
where each component of $\vec M$ is an element of the Lie algebra $\mathfrak{so}(6-n,6-n)$.  With this ansatz, the conditions above become
\be
\label{MInvariance}
h_fM_ih_f^{-1}=\lp h_b\rp^j\vphantom{\lp h_b\rp}_iM_j,\qquad\forall i,\quad\forall h\in\widehat\G,
\ee
\be
\label{MBianchi}
\ls M_i,M_j\rs=0,\qquad\forall i,j,
\ee
and a quantization condition
\be
\label{MQuantization}
\exp\ls\lambda^iM_i\rs\in\SO(6,6;\Z),\qquad\forall\vec\lambda\in\Z^6,
\ee
which can in general be a bit subtle if our base-fiber splitting is not a good splitting of the lattice.  Even in those cases however, the correct quantization condition can be worked out without too much trouble.  In the simpler case where the splitting does respect the lattice identifications, the quantization condition is simply that
\be
\exp M_i\in\SO(6-n,6-n;\Z),\qquad\forall i.
\ee

To understand how the matrices $M_i$ translate into general NS-NS flux components, we will consider R-R fields which can be thought of as sections of the spin bundle $\operatorname{Spin}(6-n,6-n)$.  The map $g(\vec x_b)$ tells us how to transport sections of this bundle as we move around the base, providing us with the correct globally defined R-R fields.  For instance, in the case with only $H$-flux, where $g(\vec x_b)$ consists solely of linear shifts in the $B$-field, we saw that the globally defined R-R-fluxes are given by $F_{RR}=\exp(B)\w F_{RR}^{(0)}$.  In this case we have
\be
dF_{RR}=\exp(B)\w\lp dF_{RR}^{(0)}+H\w F_{RR}\rp=\exp(B)\w d_HF_{RR}^{(0)}.
\ee
In other words, $d_H$ is a covariant derivative for this bundle~\cite{Micu:2007rd}, and by differentiating our globally defined sections we can deduce the form of $d_H$ and hence the components of $H$-flux.  We will now show that the same story is true more generally.  The globally defined R-R-fluxes are given by $g(\vec x_b)\cdot F_{RR}^{(0)}$, and
\be
dF_{RR}=g(\vec x_b)\cdot\mathcal{D}F_{RR}^{(0)}.
\ee
This observation is what allows us to compute the flux components from the $M_i$.

To get a better feeling for these matters, let's look at some basic cases.  Note that a general matrix in $\mathfrak{so}(6-n,6-n)$ has the form
\be
M=\lp\begin{matrix}-A^T & B \\ C & A\end{matrix}\rp,
\ee
where $A$ is a general $(6-n)\times(6-n)$ matrix, and $B$ and $C$ are antisymmetric $(6-n)\times(6-n)$ matrices.

Suppose first that all of our $M_i$ are nonvanishing only in the top-right block $B$ above, say\footnote{We now start using conventions where $i$, $j$, etc. refer to base coordinates, while $a$, $b$, etc. refer to fiber coordinates.}
\be
M_i=\lp\begin{matrix}0 & \lp B_i\rp_{ab} \\ 0 & 0\end{matrix}\rp\qquad\Longrightarrow\qquad g(\vec x_b)=\lp\begin{matrix}\mathbf 1 & \vec x_b\cdot\vec B \\ 0 & \mathbf 1\end{matrix}\rp.
\ee
Then
\be
dF_{RR}=d\lp\exp\ls\vec x_b\cdot\vec B\rs\w F_{RR}^{(0)}\rp=\exp\ls\vec x_b\cdot\vec B\rs\w\lp dF_{RR}^{(0)}+\hlf\lp B_i\rp_{ab}dx^i\w dx^a\w dx^b\w F_{RR}^{(0)}\rp,
\ee
so
\be
H_{iab}=\lp B_i\rp_{ab}.
\ee
Note that for a given base-fiber splitting we can only obtain $H$-flux with precisely one leg on the base.

Similarly, suppose that the $M_i$ are all block diagonal,
\be
M_i=\lp\begin{matrix}-A_i^T & 0 \\ 0 & A_i\end{matrix}\rp\qquad\Longrightarrow\qquad g(\vec x_b)=\lp\begin{matrix}e^{-\vec x_b\cdot\vec A^T} & 0 \\ 0 & e^{\vec x_b\cdot\vec A}\end{matrix}\rp.
\ee
Then
\bea
dF_{RR} &=& d\lp\exp\ls\hlf\Tr\lp\vec x_b\cdot\vec A\rp\rs\lp e^{-\vec x_b\cdot\vec A}\rp^\ast F_{RR}^{(0)}\rp\\
&=& e^{\hlf\Tr\lp\vec x_b\cdot\vec A\rp}\lp e^{-\vec x_b\cdot\vec A}\rp^\ast\lp dF_{RR}^{(0)}+\hlf\Tr\lp\vec A\rp\cdot d\vec x_b\w F_{RR}^{(0)}-dx^i\w\lp A_i\cdot F_{RR}^{(0)}\rp\rp,\non
\eea
where $A_i$ acts on a $p$-form via
\be
A_i\cdot \zeta^{(p)}=\binom{p}{1}\lp A_i\rp^b\vphantom{\lp A_i\rp}_{[a_1}\zeta_{|b|a_2\cdots a_p]}\,\frac{1}{p!}dx^{a_1}\w\cdots dx^{a_p}.
\ee
Comparing with (\ref{GeneralD})\footnote{In doing such a comparison, we may assume that the components of $F_{RR}^{(0)}$ are constant, so $dF_{RR}^{(0)}=0$.}, we deduce that
\be
\om^a_{ib}=\lp A_i\rp^a\vphantom{\lp A_i\rp}_b.
\ee
Again we find that $\om$ must have exactly one lower index along the base, with the other two indices along the fiber.  Note that here we do not require $A_i$ to be traceless, though any nonvanishing trace piece would require a base one-form $\Tr(A_i)dx^i$ which would have to be invariant under the orientifold group.

And also,
\be
M_i=\lp\begin{matrix}0 & 0 \\ \lp C_i\rp^{ab} & 0\end{matrix}\rp\qquad\Longrightarrow\qquad g(\vec x_b)=\lp\begin{matrix}\mathbf 1 & 0 \\ \vec x_b\cdot\vec C & \mathbf 1\end{matrix}\rp.
\ee
So from
\be
dF_{RR}=d\lp\exp\ls\hlf x^iC_i^{ab}\iota_a\iota_b\rs\cdot F_{RR}^{(0)}\rp=\exp\ls\hlf x^iC_i^{ab}\iota_a\iota_b\rs\cdot\lp dF_{RR}^{(0)}+\hlf C_i^{ab}dx^i\w\lp\iota_a\iota_b F_{RR}^{(0)}\rp\rp,
\ee
we find
\be
Q^{ab}_i=-\lp C_i\rp^{ab}.
\ee
Once again, the lower index must be on the base, while the other two (upper) indices lie along the fiber.

Finally, since the exponent of $g$ is linear in the base coordinates, these derivatives simply add, and we find that the map between the matrices $M_i$ and the fluxes is simply,
\be
\label{MiFluxes}
M_i=\lp\begin{matrix}-\om^b_{ia} & H_{iab} \\ -Q^{ab}_i & \om^a_{ib}\end{matrix}\rp.
\ee

Let us see what we can learn from the constraints (\ref{MInvariance}) and (\ref{MBianchi}).  Consider an element of $\widehat\G$ of the form
\be
h=\Om\hat\s=\lp\begin{matrix}-\lp\s^T\rp^{-1} & 0 \\ 0 & \s\end{matrix}\rp=\s_b\oplus\lp\begin{matrix}-\lp\s_f^T\rp^{-1} & 0 \\ 0 & \s_f\end{matrix}\rp.
\ee
Then substituting (\ref{MiFluxes}) into (\ref{MInvariance}) leads to
\be
-\lp\s_f^T\rp^{-1}H_i\s_f^{-1}=\lp\s_b\rp^j\vphantom{\lp\s_b\rp}_iH_j,\qquad\s_fQ_i\s_f^T=-\lp\s_b\rp^j\vphantom{\lp\s_b\rp}_iQ_j,\qquad\s_f\om_i\s_f^{-1}=\lp\s_b\rp^j\vphantom{\lp\s_b\rp}_i\om_j,
\ee
which can be rephrased as the statement that the metric fluxes $\om$ should be even under the involution $\s$, while the $H$- and $Q$-fluxes should both be odd under $\s$.

Now substituting (\ref{MiFluxes}) into (\ref{MBianchi}) leads to the conditions
\be
\om^a_{c[i}\om^c_{j]b}+Q^{ac}_{[i}H_{j]cb}=0,\qquad H_{ac[i}\om^c_{j]b}-H_{bc[i}\om^c_{j]a}=0,\qquad Q^{c[a}_{[i}\om^{b]}_{j]c}=0.
\ee
But it is easy to check that these are precisely the Bianchi identities (\ref{AppendixBI}) for the situation at hand, namely when each flux has exactly one lower index on the base and all other indices lie along the fiber.

We would like to discuss the quantization condition (\ref{MQuantization}), but it is quite complicated in the general case, so let us first see how these base-fiber constructions work in our favorite example.

\subsection{Example}
\label{BFExample}

To classify the possible base-fiber splittings of our $T^6/\Z_4$ orientifold, we need to know how the coordinates of the $T^6$ split into representations of the orientifold group $D_4$.  As a real vector space (i.e. forgetting the shift identifications of the torus), it can be checked that this $\R^6$ splits into two isomorphic two-dimensional irreducible real representations and two one-dimensional real representations which are not isomorphic.  The latter two are given by the span of $y_3$ and the span of $\hat x_3=x_3+\hlf y_3$.  Because of the isomorphism between the two-dimensional representations, there is a two real parameter family of ways to split up the first four coordinates into irreducible real representations.  Indeed, if we define
\be
\label{BasisTransformation}
\begin{array}{ll}
\hat x_1=x_1+a\lp x_2+y_2\rp,\qquad & \hat y_1=y_1+a\lp-x_2+y_2\rp,\qquad \\ \hat x_2=b\lp x_1-y_1\rp+x_2,\qquad & \hat y_2=b\lp x_1+y_1\rp+y_2,
\end{array}
\ee
then $\{\hat x_1,\hat y_1\}$ can be taken to span one invariant subspace, while $\{\hat x_2,\hat y_2\}$ span the other.  The only constraint is that $2ab\ne 1$, so that this change of basis is invertible.

We can now classify all of the possible bases, dimension by dimension.

\subsubsection{One-dimensional bases}

Here there are two cases; either the base is parametrized by $y_3$, or by $\hat x_3=x_3+\hlf y_3$.  Suppose that the base is $y_3$.  Invariance under $\Theta^2$ ensures that $M_{y_3}$ has the form
\be
M_{y_3}=\lp\begin{matrix}A & 0 & B & 0 \\ 0 & a & 0 & 0 \\ C & 0 & -A^T & 0 \\ 0 & 0 & 0 & -a\end{matrix}\rp,
\ee
for $4\times 4$ matrices $A$, $B$, $C$, and a real number $a$.  But now invariance under $\Theta$ implies that $a=0$, and that $M_{y_3}$ in fact lies in $\mathfrak{so}(4,4)$.  But then this one-dimensional case is really a restriction of the case with two-dimensional base $T^2_3$ where only $y_3$ dependence is allowed (i.e. $M_{\hat x_3}=0$).  This case is treated below without restriction.

Since we did not use the action of $\s$ in the argument above, and since this action is the only difference between $y_3$ and $\hat x_3$, we conclude that an $\hat x_3$ base also gives nothing new.

\subsubsection{Two-dimensional bases}

Here we consider four-dimensional fibers and two-dimensional bases.  There are three possibilities.

\vskip 0.2cm
1)\quad{\underline{$T^2_1$ base}}

In this case the base is spanned by $\{\hat x_1,\hat y_1\}$, and the fiber is spanned by $\{\hat x_2,\hat y_2,x_3,y_3\}$.  Our orientifold group is represented as
\be
\Theta_b=\lp\begin{matrix}0 & -1 \\ 1 & 0\end{matrix}\rp,\qquad\s_b=\lp\begin{matrix}1 & 0 \\ 0 & -1\end{matrix}\rp,
\ee
\be
\hat\Theta_f=\lp\begin{matrix}\begin{smallmatrix}0 & -1 & 0 & 0 \\ 1 & 0 & 0 & 0 \\ 0 & 0 & -1 & 0 \\ 0 & 0 & 0 & -1\end{smallmatrix} & 0 \\ 0 & \begin{smallmatrix}0 & -1 & 0 & 0 \\ 1 & 0 & 0 & 0 \\ 0 & 0 & -1 & 0 \\ 0 & 0 & 0 & -1\end{smallmatrix} \end{matrix}\rp,\qquad\lp\Om\hat\s\rp_f=\lp\begin{matrix}\begin{smallmatrix}0 & -1 & 0 & 0 \\ -1 & 0 & 0 & 0 \\ 0 & 0 & -1 & 0 \\ 0 & 0 & -1 & 1\end{smallmatrix} & 0 \\ 0 & \begin{smallmatrix}0 & 1 & 0 & 0 \\ 1 & 0 & 0 & 0 \\ 0 & 0 & 1 & 1 \\ 0 & 0 & 0 & -1\end{smallmatrix}\end{matrix}\rp.
\ee

The constraints (\ref{MInvariance}) imply that
\be
\hat\Theta_f^2M_{\hat x_1}\hat\Theta_f^2=-M_{\hat x_1}\qquad\Longrightarrow\qquad M_{\hat x_1}=\lp\begin{matrix}0 & a & 0 & c \\ b & 0 & -c^T & 0 \\ 0 & d & 0 & -b^T \\ -d^T & 0 & -a^T & 0\end{matrix}\rp,
\ee
where $a$, $b$, $c$, and $d$ are two-by-two matrices.  Next,
\be
\label{Mhatx1}
\lp\Om\hat\s\rp_fM_{\hat x_1}\lp\Om\hat\s\rp_f=M_{\hat x_1}\qquad\Longrightarrow\qquad M_{\hat x_1}=\lp\begin{matrix}0 & \begin{smallmatrix}\alpha & \beta \\ \alpha+\beta & -\beta\end{smallmatrix} & 0 & \begin{smallmatrix}\varepsilon & \varphi \\ -\varepsilon & \varphi-\varepsilon\end{smallmatrix} \\ \begin{smallmatrix}\g & \g \\ \d & \g-\d\end{smallmatrix} & 0 & \begin{smallmatrix}-\varepsilon & \varepsilon \\ -\varphi & \varepsilon-\varphi\end{smallmatrix} & 0 \\ 0 & \begin{smallmatrix}\chi & \k \\ -\chi-\k & \k\end{smallmatrix} & 0 & \begin{smallmatrix}-\g & -\d \\ -\g & \d-\g\end{smallmatrix} \\ \begin{smallmatrix}-\chi & \chi+\k \\ -\k & -\k\end{smallmatrix} & 0 & \begin{smallmatrix}-\alpha & -\alpha-\beta \\ -\beta & \beta\end{smallmatrix} & 0\end{matrix}\rp.
\ee
Then all of the constraints (\ref{MInvariance}) are satisfied if we define
\be
\label{Mhaty1}
M_{\hat y_1}=\hat\Theta_fM_{\hat x_1}\hat\Theta_f^{-1}=\lp\begin{matrix}0 & \begin{smallmatrix}\al+\beta & -\beta \\ -\al & -\beta\end{smallmatrix} & 0 & \begin{smallmatrix}-\varepsilon & \varphi-\varepsilon \\ -\varepsilon & -\varphi\end{smallmatrix} \\ \begin{smallmatrix}\g & -\g \\ \g-\d & -\d\end{smallmatrix} & 0 & \begin{smallmatrix}\varepsilon & \varepsilon \\ \varepsilon-\varphi & \varphi\end{smallmatrix} & 0 \\ 0 & \begin{smallmatrix}-\chi-\k & \k \\ -\chi & -\k\end{smallmatrix} & 0 & \begin{smallmatrix}-\g & \d-\g \\ \g & \d\end{smallmatrix} \\ \begin{smallmatrix}\chi+\k & \chi \\ -\k & \k\end{smallmatrix} & 0 & \begin{smallmatrix}-\al-\beta & \al \\ \beta & \beta\end{smallmatrix} & 0\end{matrix}\rp.
\ee

By imposing the requirement that these matrices commute, we find three extra conditions, namely
\be
\label{T21BI}
\beta\g+\varepsilon\k=0,\qquad\al\g+\beta\d=\varepsilon\chi+\varphi\k,\qquad\lp\al+\beta\rp\d+\lp\varphi-\varepsilon\rp\chi=0.
\ee

From the entries of $M_{\hat x_1}$ and $M_{\hat y_1}$ one can read off the flux components in that basis.  One then uses the transformation (\ref{BasisTransformation}) to convert these fluxes back to the lattice compatible basis from before.  The resulting fluxes are
\be
p=\Delta\lp\begin{matrix}\varphi-\varepsilon & \varphi\end{matrix}\rp,\non
\ee
\be
r=\lp\begin{matrix}2a^2\Delta^{-1}\d & 2a^2\Delta^{-1}\lp\d-\g\rp \\ \Delta^{-1}\d & \Delta^{-1}\lp\d-\g\rp \\ -\Delta\lp\al+\beta\rp & -\Delta\al \\ 4a\Delta^{-1}\d & 4a\Delta^{-1}\lp\d-\g\rp\end{matrix}\rp,\qquad q=\Delta^{-1}\lp\begin{matrix}\chi & \chi+\k \\ 2a^2\chi & 2a^2\lp\chi+\k\rp \\ 0 & 0 \\ 4a\chi & 4a\lp\chi+\k\rp\end{matrix}\rp,
\ee
with $\widehat r=\widehat q=s=0$, and where $a$ and $\Delta=1-2ab$ are the parameters of the basis transformation.  With these definitions, one can check that the constraints (\ref{T21BI}) precisely reproduce the Bianchi identities (\ref{ExampleBI}) for this case.

Now unless $a$, $b$, and $\Delta$ are integers, the basis in which the matrices above are expressed is not a basis for our lattice, and so generally the quantization condition is not just that $\exp[M_{\hat x_1}]$ and $\exp[M_{\hat y_1}]$ are integers.  Instead, what we should do is embed these matrices into $\mathfrak{so}(6,6)$, undo the transformation (\ref{BasisTransformation}), and then exponentiate.  Following this procedure we find $12\times 12$ matrices $M_1$, $M_2$, as well as $M_3=a(M_1-M_2)$ and $M_4=a(M_1+M_2)$.  All four of these matrices turn out to be (three-step) nilpotent, and hence the quantization conditions $\exp[M_i]\in\SO(6,6;\Z)$ are simply that the entries of the $M_i$ be integers.  Translating back into the matrices above, we learn that the correct quantization condition for this case is nearly the naive one (in fact it is the naive quantization condition in terms of the flux components, $\om^i_{jk}$, $q^{ij}_k$, etc.); we must have $p_K$, $r_{cK}$ and $q_{cK}$ to be integers, but in addition we require $2ar_{3K}$, $2ap_K$, $a(r_{31}-r_{32})$, and $a(p_1-p_2)$ to be integers.  In particular, if $a$ is an integer (for instance if the transformed basis is a lattice basis), then the naive integer quantization is correct.

Let us take a moment and consider the types of solutions that we get if we restrict to the case $q=0$ ($\chi=\k=0$).  Then the Bianchi identities (\ref{T21BI}) force either $\al=\beta=0$, or $\g=\d=0$.  Either way, we are stuck with an $r$-matrix of rank one, and, following our discussion in section~\ref{omegaExample}, we cannot stabilize all of the moduli.

\vskip 0.2cm
2)\quad{\underline{$T^2_2$ base}}

Here we take our base to be spanned by $\{\hat x_2,\hat y_2\}$.  This case works out almost identically to the case described in detail above.  In fact, the expression for $M_{\hat x_2}$ is precisely the same as that for $M_{\hat x_1}$ in (\ref{Mhatx1}), while $M_{\hat y_2}$ is the same as $M_{\hat y_1}$ in (\ref{Mhaty1}).  As such, the Bianchi identities are again simply the three equations in (\ref{T21BI}).  What does change slightly is the map back to our flux matrices.  For this case we have
\be
p=\Delta\lp\begin{matrix}\varepsilon-\varphi & -\varphi\end{matrix}\rp,\non
\ee
\be
r=\lp\begin{matrix}-\Delta^{-1}\d & \Delta^{-1}\lp\g-\d\rp \\ -2b^2\Delta^{-1}\d & 2b^2\Delta^{-1}\lp\g-\d\rp \\ \Delta\lp\al+\beta\rp & \Delta\al \\ -4b\Delta^{-1}\d & 4b\Delta^{-1}\lp\g-\d\rp\end{matrix}\rp,\qquad q=-\Delta^{-1}\lp\begin{matrix}2b^2\chi & 2b^2\lp\chi+\k\rp \\ \chi & \chi+\k \\ 0 & 0 \\ 4b\chi & 4b\lp\chi+\k\rp\end{matrix}\rp,
\ee
and $\widehat r=\widehat q=s=0$.  The quantization conditions are exactly as before but with $a$ replaced by $b$ wherever it occurs.

\vskip 0.2cm
3)\quad{\underline{$T^2_3$ base}}

Finally there is the case in which our base is spanned by $\{x_3,y_3\}$.  This case turns out to be richer than the previous cases.  The representation of the orientifold group is
\be
\Theta_b=\lp\begin{matrix}-1 & 0 \\ 0 & -1\end{matrix}\rp,\qquad\s_b=\lp\begin{matrix}1 & 1 \\ 0 & -1\end{matrix}\rp,
\ee
\be
\hat\Theta_f=\lp\begin{matrix}\begin{smallmatrix}0 & -1 & 0 & 0 \\ 1 & 0 & 0 & 0 \\ 0 & 0 & 0 & -1 \\ 0 & 0 & 1 & 0\end{smallmatrix} & 0 \\ 0 & \begin{smallmatrix}0 & -1 & 0 & 0 \\ 1 & 0 & 0 & 0 \\ 0 & 0 & 0 & -1 \\ 0 & 0 & 1 & 0\end{smallmatrix}\end{matrix}\rp,\qquad\lp\Om\hat\s\rp_f=\lp\begin{matrix}\begin{smallmatrix}-1 & 0 & 0 & 0 \\ 0 & 1 & 0 & 0 \\ 0 & 0 & 0 & -1 \\ 0 & 0 & -1 & 0\end{smallmatrix} & 0 \\ 0 & \begin{smallmatrix}1 & 0 & 0 & 0 \\ 0 & -1 & 0 & 0 \\ 0 & 0 & 0 & 1 \\ 0 & 0 & 1 & 0\end{smallmatrix}\end{matrix}\rp.
\ee
Solving the constraints
\be
\begin{array}{ll}\hat\Theta_fM_{x_3}\hat\Theta_f^{-1}=-M_{x_3}, & \lp\Om\hat\s\rp_fM_{x_3}\lp\Om\hat\s\rp_f=M_{x_3}, \\ \hat\Theta_fM_{y_3}\hat\Theta_f^{-1}=-M_{y_3}, & \lp\Om\hat\s\rp_fM_{y_3}\lp\Om\hat\s\rp_f=M_{x_3}-M_{y_3},\end{array}
\ee
leads to a twelve-parameter family of solutions,
\bea
M_{x_3} &=& \lp\begin{matrix}\begin{smallmatrix}\al & 0 \\ 0 & -\al\end{smallmatrix} & \begin{smallmatrix}\g & \g \\ \g & -\g\end{smallmatrix} & 0 & \begin{smallmatrix}\varepsilon & -\varepsilon \\ -\varepsilon & -\varepsilon\end{smallmatrix} \\ \begin{smallmatrix}\beta & \beta \\ \beta & -\beta\end{smallmatrix} & \begin{smallmatrix}0 & \d \\ \d & 0\end{smallmatrix} & \begin{smallmatrix}-\varepsilon & \varepsilon \\ \varepsilon & \varepsilon\end{smallmatrix} & 0 \\ 0 & \begin{smallmatrix}\n & -\n \\ -\n & -\n\end{smallmatrix} & \begin{smallmatrix}-\al & 0 \\ 0 & \al\end{smallmatrix} & \begin{smallmatrix}-\beta & -\beta \\ -\beta & \beta\end{smallmatrix} \\ \begin{smallmatrix}-\n & \n \\ \n & \n\end{smallmatrix} & 0 & \begin{smallmatrix}-\g & -\g \\ -\g & \g\end{smallmatrix} & \begin{smallmatrix}0 & -\d \\ -\d & 0\end{smallmatrix}\end{matrix}\rp,\\
M_{y_3} &=& \lp\begin{matrix}\begin{smallmatrix}\al/2 & \varphi \\ \varphi & -\al/2\end{smallmatrix} & \begin{smallmatrix}\k+\g/2 & -\k+\g/2 \\ -\k+\g/2 & -\k-\g/2\end{smallmatrix} & 0 & \begin{smallmatrix}\m+\varepsilon/2 & \m-\varepsilon/2 \\ \m-\varepsilon/2 & -\m-\varepsilon/2\end{smallmatrix} \\ \begin{smallmatrix}\chi+\beta/2 & -\chi+\beta/2 \\ -\chi+\beta/2 & -\chi-\beta/2\end{smallmatrix} & \begin{smallmatrix}\lambda & \d/2 \\ \d/2 & -\lambda\end{smallmatrix} & \begin{smallmatrix}-\m-\varepsilon/2 & -\m+\varepsilon/2 \\ -\m+\varepsilon/2 & \m+\varepsilon/2\end{smallmatrix} & 0 \\ 0 & \begin{smallmatrix}\pi+\n/2 & \pi-\n/2 \\ \pi-\n/2 & -\pi-\n/2\end{smallmatrix} & \begin{smallmatrix}-\al/2 & -\varphi \\ -\varphi & \al/2\end{smallmatrix} & \begin{smallmatrix}-\chi-\beta/2 & \chi-\beta/2 \\ \chi-\beta/2 & \chi+\beta/2\end{smallmatrix} \\ \begin{smallmatrix}-\pi-\n/2 & -\pi+\n/2 \\ -\pi+\n/2 & \pi+\n/2\end{smallmatrix} & 0 & \begin{smallmatrix}-\k-\g/2 & \k-\g/2 \\ \k-\g/2 & \k+\g/2\end{smallmatrix} & \begin{smallmatrix}-\lambda & -\d/2 \\ -\d/2 & \lambda\end{smallmatrix}\end{matrix}\rp.\non
\eea
Note that the solution is much simpler in terms of the matrix $M_{\hat y_3}=M_{y_3}-\hlf M_{x_3}$, but that the solution as given corresponds to the basis for the lattice.

Enforcing $[M_{x_3},M_{y_3}]=0$ gives six equations
\be
\lp\al+\d\rp\chi-\lp\varphi-\lambda\rp\beta=\lp\al+\d\rp\pi+\lp\varphi-\lambda\rp\n=0,\non
\ee
\be
\label{T23BI}
\lp\al+\d\rp\k-\lp\varphi-\lambda\rp\g=\lp\al+\d\rp\m+\lp\varphi-\lambda\rp\varepsilon=0,
\ee
\be
\al\varphi-\g\chi-\beta\k-\m\n-\varepsilon\pi=0,\qquad\al\varphi+\d\lambda=0.\non
\ee

Translating into flux matrices, we find
\be
p=\lp\begin{matrix}\m-\varepsilon/2 & \m+\varepsilon/2\end{matrix}\rp,
\ee
\be
r=\lp\begin{matrix}\chi+\beta/2 & \chi-\beta/2 \\ -\k-\g/2 & -\k+\g/2 \\ 0 & 0 \\ -\varphi-\lambda+\hlf(\al-\d) & -\varphi-\lambda-\hlf(\al-\d)\end{matrix}\rp,\qquad q=\lp\begin{matrix}0 & 0 \\ 0 & 0 \\ -\pi+\n/2 & -\pi-\n/2 \\ 0 & 0\end{matrix}\rp,\non
\ee
\be
\widehat r=\lp\begin{matrix}\varphi-\lambda-\hlf(\al+\d) & -\varphi+\lambda-\hlf(\al+\d)\end{matrix}\rp,\qquad\widehat q=s=0.\non
\ee
In the case $q=0$ ($\pi=\n=0$), this provides the complete case $(b)$ of section~\ref{omegaExample}.

One can easily verify that (\ref{T23BI}) gives the correct set of Bianchi identities for this case.  In fact the solution to these equations can be broken into four cases,
\be
\begin{array}{cl}(\mathrm{i}) & M_{x_3}=0, \\ (\mathrm{ii}) & M_{y_3}=\hlf M_{x_3}, \\ (\mathrm{iii}) & \al+\d=\varphi-\lambda=0, \\ & \al\varphi-\g\chi-\beta\k-\m\n-\varepsilon\pi=0, \\ (\mathrm{iv}) & \al+\d\ne 0,\ \varphi-\lambda\ne 0, \\ & \chi=\lp\frac{\varphi-\lambda}{\al+\d}\rp\beta,\ \pi=-\lp\frac{\varphi-\lambda}{\al+\d}\rp\n,\ \k=\lp\frac{\varphi-\lambda}{\al+\d}\rp\g,\ \m=-\lp\frac{\varphi-\lambda}{\al+\d}\rp\varepsilon, \\ & \al-\d=\pm\ls\lp\al+\d\rp^2-8\beta\g+8\varepsilon\n\rs^{1/2},\ \varphi+\lambda=-\lp\frac{\varphi-\lambda}{\al+\d}\rp\lp\al-\d\rp.\end{array}
\ee
In terms of flux matrices, case $(\mathrm{i})$ has $p_1=p_2$, $r_{a1}=r_{a2}$, $q_{31}=q_{32}$, and $\widehat r_1=-\widehat r_2$.  Case $(\mathrm{ii})$ corresponds to $p_1=-p_2$, $r_{a1}=-r_{a2}$, $q_{31}=-q_{32}$, and $\widehat r_1=\widehat r_2$.  Case $(\mathrm{iii})$ is simply $\widehat r=0$, with the other components arbitrary (up to one additional Bianchi identity).  And case $(\mathrm{iv})$ is the case with arbitrary $\widehat r$, but where the conditions $\widehat r_Kp_K=\widehat r_Kr_{aK}=\widehat r_Kq_{3K}=0$ put constraints on the other fluxes.

In every case we must finally solve the quantization conditions $\exp[M_{x_3}],\exp[M_{y_3}]\in\SO(4,4;\Z)$.  In each of the four cases this condition is potentially nontrivial because at least one of the two matrices may not be nilpotent.  For example, the general expression for the exponentiated version of $M_{x_3}$ includes entries such as
\be
e^{\hlf\lp\al+\d\rp}\ls\cosh\frac{C}{2}+\frac{\al-\d}{C}\sinh\frac{C}{2}\rs,\quad\mathrm{or}\quad\frac{2\beta}{C}e^{\hlf\lp\al+\d\rp}\sinh\frac{C}{2},
\ee
and many others, where
\be
C=\sqrt{\lp\al-\d\rp^2+8\beta\g-8\varepsilon\n}.
\ee
Finding the generic situation in which all of these entries are integers is quite difficult.  Let us specialize somewhat.

To make contact with the work we did in section~\ref{omegaExample}, we will focus on case $(\mathrm{iii})$, where $\widehat r=0$, and assume also that $q=0$.  Under what conditions would the naive, integral quantization be correct?  The requirement would be that both $M_{x_3}$ and $M_{y_3}$ would have to be nilpotent, and this in turn requires that two expressions vanish,
\be
\al^2+2\beta\g=0,\qquad\varphi^2+2\k\chi=0.
\ee
And it turns out that these equations, along with the extra Bianchi identity $\al\varphi-\beta\k-\g\chi=0$, imply that the rank of $r$ is one.  Thus, by arguments in section~\ref{EFTApproach}, we cannot hope to stabilize all moduli.  In particular, the numerical solution we presented at the end of section~\ref{omegaExample} is not correctly quantized.  In fact, it is possible to find solutions to the quantization conditions which do give rise to an $r$-matrix of rank two and a superpotential which stabilizes all moduli.  However, we have argued that such cases are not nilpotent, so the entries of the $r$-matrix are not integers and in fact are irrational numbers.  But now we have a puzzle, since if all the NS-NS fluxes are irrational numbers, then it is clearly impossible to satisfy the tadpole condition with R-R flux integers!

One plausible solution is that we do not correctly understand the quantization of R-R fluxes in the presence of general NS-NS fluxes, and in particular in non-nilpotent cases where the NS-NS flux quantization is not the naive one.  One approach to this problem would involve viewing both NS-NS and R-R fluxes as twists in a U-duality group of the fiber, in which case understanding the full quantization conditions would simply reduce to understanding the structure of the duality group, e.g. $E_{7(7)}(\Z)$.  This is an avenue of ongoing investigation.

\subsubsection{Three-dimensional bases}

There are four possible bases in this case, but it will turn out that they are all contained in previously considered examples, so we will focus just on the case with base $\{\hat x_1,\hat y_1,\hat x_3\}$.  The other three cases (with either or both of $\{\hat x_2,\hat y_2\}$ or $\hat y_3$) are similar.

Here we have
\be
\Theta_b=\lp\begin{matrix}0 & -1 & 0 \\ 1 & 0 & 0 \\ 0 & 0 & -1\end{matrix}\rp,\qquad\s_b=\lp\begin{matrix}1 & 0 & 0 \\ 0 & -1 & 0 \\ 0 & 0 & 1\end{matrix}\rp,
\ee
\be
\hat\Theta_f=\lp\begin{matrix}\begin{smallmatrix}0 & -1 \\ 1 & 0\end{smallmatrix} & & & \\ & -1 & & \\ & & \begin{smallmatrix}0 & -1 \\ 1 & 0\end{smallmatrix} & \\ & & & -1\end{matrix}\rp,\qquad\lp\Om\hat\s\rp_f=\lp\begin{matrix}\begin{smallmatrix}0 & -1 \\ -1 & 0\end{smallmatrix} & & & \\ & 1 & & \\ & & \begin{smallmatrix}0 & 1 \\ 1 & 0\end{smallmatrix} & \\ & & & -1\end{matrix}\rp.
\ee

Solving our constraints, we find
\be
M_{\hat x_1}=\lp\begin{matrix}0 & \begin{smallmatrix}\g \\ -\g\end{smallmatrix} & 0 & \begin{smallmatrix}-\d \\ -\d\end{smallmatrix} \\ \begin{smallmatrix}\al & -\al\end{smallmatrix} & 0 & \begin{smallmatrix}\d & \d\end{smallmatrix} & 0 \\ 0 & \begin{smallmatrix}-\beta \\ -\beta\end{smallmatrix} & 0 & \begin{smallmatrix}-\al \\ \al\end{smallmatrix} \\ \begin{smallmatrix}\beta & \beta\end{smallmatrix} & 0 & \begin{smallmatrix}-\g & \g\end{smallmatrix} & 0\end{matrix}\rp,\qquad M_{\hat y_1}=\lp\begin{matrix}0 & \begin{smallmatrix}-\g \\ -\g\end{smallmatrix} & 0 & \begin{smallmatrix}-\d \\ \d\end{smallmatrix} \\ \begin{smallmatrix}-\al & -\al\end{smallmatrix} & 0 & \begin{smallmatrix}\d & -\d\end{smallmatrix} & 0 \\ 0 & \begin{smallmatrix}-\beta \\ \beta\end{smallmatrix} & 0 & \begin{smallmatrix}\al \\ \al\end{smallmatrix} \\ \begin{smallmatrix}\beta & -\beta\end{smallmatrix} & 0 & \begin{smallmatrix}\g & \g\end{smallmatrix} & 0\end{matrix}\rp,\non
\ee
\be
M_{\hat x_3}=\lp\begin{matrix}\begin{smallmatrix}0 & \varepsilon \\ \varepsilon & 0\end{smallmatrix} & & & \\ & 0 & & \\ & & \begin{smallmatrix}0 & -\varepsilon \\ -\varepsilon & 0\end{smallmatrix} & \\ & & & 0\end{matrix}\rp.
\ee

Now let us check the Bianchi identities.  It turns out that they give
\be
\al\g=\beta\d=0,\quad\mathrm{and}\quad\varepsilon\al=\varepsilon\beta=\varepsilon\g=\varepsilon\d=0.
\ee
But these then imply that either $M_{\hat x_3}=0$, and we are in a special case of two-dimensional bases, or that $M_{\hat x_1}=M_{\hat y_1}=0$, and we are in a special case of a one-dimensional base.  Either way, we find nothing new.

The other three possible bases lead to the same conclusions.

\subsubsection{Four-dimensional bases}

Here there are three possibilities.  If our base is given by $\{\hat x_1,\hat y_1,x_3,y_3\}$, then we would have
\be
(-{\bf 1})M_{\hat x_1}(-{\bf 1})=\hat\Theta_f^2M_{\hat x_1}\hat\Theta_f^2=\lp\Theta_b^2\cdot M\rp_{\hat x_1}=-M_{\hat x_1}\qquad\Longrightarrow\qquad M_{\hat x_1}=0,
\ee
and similaraly $M_{\hat y_1}=0$, so our base is equivalent to just having $\{x_3,y_3\}$.

By the same argument, a base of $\{\hat x_2,\hat y_2,x_3,y_3\}$ would reduce to a previously considered case.  This leaves only the possibility of $\{x_1,y_1,x_2,y_2\}$.  But then we have, e.g.
\be
({\bf 1})M_{x_1}({\bf 1})=\hat\Theta_f^2M_{x_1}\hat\Theta_f^2=-M_{x_1},
\ee
and this case in fact forces all $M_i=0$.

\subsubsection{Five-dimensional bases}

This final case is also trivial for the same reasons discussed above.  Invariance under $\Theta^2$ forces four of the $M_i$ to vanish, and invariance under $\Theta$ takes care of the fifth one.

\subsection{Advantages and Puzzles}
\label{AdvantagesAndPuzzles}

In this section we have presented a ten-dimensional construction of IIA toroidal orientifold models with some general NS-NS fluxes.  The class of models which we can construct in this way is a sub-class of all the models discussed in section~\ref{EFTApproach}.  It is not clear how representative a sample this sub-class is.  For instance, in the case with only metric flux, the models we can construct do not necessarily correspond to nilpotent algebras, and hence are more general than those obtained by T-dualizing $H$-flux alone, but they are still a restricted set of algebras, and in particular are all solvable algebras.  It would be interesting to compare these properties with the geometric properties evidenced, for example, in the classification of twisted tori in~\cite{Grana:2006kf}.  For our $T^6/\Z_4$ example we can construct nearly all possible metric fluxes (all of cases $(a)$ and $(b)$, but not the cases $(a')$ or $(a'')$), but if non-geometric fluxes are included then only a fairly small fraction of possible models can be built in this way.

Having any kind of ten-dimensional construction, however, is obviously a huge advantage, as it gives us a great deal more confidence that our models can really arise from string theory compactifications (though we are certainly not claiming that other models can't be obtained from string theory, just not using the methods we have explored in this paper).  It can also highlight important subtleties that were not so readily apparent from the effective theory approach.  As we have seen, the quantization of general NS-NS fluxes is one such subtlety.  In cases where our matrices $M_i$ are all two-step nilpotent (which also implies that the underlying Lie algebra, as described in appendix~\ref{BIDerivation}, is nilpotent) then the quantization condition is simply the naive one, with all flux components (which correspond to entries of the $M_i$) being integers.  Matrices that are nilpotent after more than two-steps must have entries which are rationals (with denominator no larger than the number of steps minus one).  More generally, however the condition is that certain exponentials of matrices be integral.  These conditions often can be solved, giving irrational flux components (see also related discussions in~\cite{Dabholkar:2002sy}).

This in turn leads to a puzzle, since the integral tadpole contribution is given by a bilinear pairing of the NS-NS fluxes with the R-R fluxes.  If the former are forced by quantization to be irrational numbers, then the latter cannot be integers or rationals, as was presumed.  Either such setups are inconsistent or (more likely, in our belief) we have not correctly understood the quantization of R-R fluxes in general NS-NS flux backgrounds (see also the discussion in footnote~\ref{RRQuantizationFootnote}).  Presumably there should be some analog of twisted K-theory for the general flat fiber theories that we are studying, including also the non-geometric fluxes.  Matching this onto the work of Mathai and collaborators~\cite{Bouwknegt:2003vb,Bouwknegt:2003wp,Bouwknegt:2003zg,Mathai:2004qq,Mathai:2004qc,Bouwknegt:2004ap,Bouwknegt:2004tr,Mathai:2005fd} would be very interesting.  Similarly, exploiting the connections between the base-fiber approach described here and spaces with generalized complex structure (see e.g. \cite{Gualtieri:2003dx}) could potentially lead to a better understanding of these more general classes of string compactifications.  We are currently investigating these directions.

One advantage to following this base-fiber approach is that the constructions should be easy to generalize to any situation with a flat fiber over a flat base (and some aspects of the approach should be applicable to more general smooth bases).  In particular, we should certainly be able to accommodate type IIB as well as IIA within the framework presented, and we can also work with orientifold actions which are asymmetric on the fiber (compare for instance with the models of~\cite{Becker:2006ks}).  Furthermore, heterotic string theory on a torus or type II on K3 fibers can also be covered in this framework, since the duality groups in those situations are well-understood.  The K3 fibered case in particular could be very interesting, and could be compared to the work of~\cite{Cvetic:2007ju}.  Finally, one can expand the analysis to include U-duality groups, such as that of M-theory on a torus (see~\cite{Kumar:1996zx,Dabholkar:2002sy,Hull:2003kr,Hull:2004in,Dabholkar:2005ve,Hull:2006tp,Hull:2007zu,Aldazabal:2006up}).  It would be very interesting to understand how far one could push such a program, and whether one could find interesting solutions with a controlled low energy theory.

We have not yet incorporated any of the twisted sector physics into this story.  Beyond getting possible hints by studying K3 fibers at their orbifold points, it would be extremely gratifying to have a more complete picture for how to deal with the twisted sector physics in the presence of fiber twists.

\section*{Acknowledgements}
It is a pleasure to thank Aaron Bergman, Jacques Distler, Raphael
Flauger, Sonia Paban and Uday Varadarajan for many helpful comments,
discussions and encouragement. The research of the authors is based
upon work supported by the National Science Foundation under Grant
No. PHY-0455649.

\appendix

\section{$SU(3)$ Structure with Metric Fluxes}
\label{SU3Structure}

In this appendix we are analyzing the $\SU(3)$ structure of the
torus with metric fluxes, as was done in a specific case in~\cite{Camara:2005dc}. For a general discussion of $\SU(3)$ structures, see
for example \cite{Grana:2005jc}.

We start with the K\"ahler 2-form $J$ and the holomorphic three-form $\Om$ of the geometry, given by
\bea
J &=& v_a\om_a,\\
\Om &=& \mathcal{Z}_Ka_K-\mathcal{F}_Kb_K,
\eea
where $\mathcal{Z}_K$ are real and $\mathcal{F}_K$ are imaginary.  These forms satisfy $J \wedge \Omega=0, \, J \wedge J \wedge J=6 \ic
\mathcal{V}_6 \Omega \wedge \overline{\Omega}$ and define an $\SU(3)$
structure on the twisted torus. The torsion classes are defined by
\begin{align}
\text{d} J &= -12 \mathcal{V}_6 \text{Im}(\mathcal{W}_1 \overline{\Omega}) + \mathcal{W}_4 \wedge J + \mathcal{W}_3,\\
\text{d} \Omega &= \mathcal{W}_1 J \wedge J + \mathcal{W}_2 \wedge J
+ \mathcal{W}_5^* \wedge \Omega.
\end{align}
$\mathcal{W}_1$ is a complex scalar, $\mathcal{W}_2$ is a complex
primitive (1,1)-form i.e., $\mathcal{W}_2 \wedge J \wedge J=0$,
$\mathcal{W}_3$ is a real primitive (2,1)+(1,2)-form i.e.,
$\mathcal{W}_3 \wedge J=\mathcal{W}_3 \wedge \Omega =0$, $\mathcal{W}_4$ is a real 1-form and
$\mathcal{W}_5$ is a complex (1,0)-form. The prefactor in the first term of d$J$ is needed to have d$(J\wedge \Omega)=0$.\\
The torsion classes can be read off from
\begin{align}
\label{eq:dJ} \text{d} J &= -r_{aK} v_a b_K,\\
\text{d} \Omega &= -\mathcal{Z}_K \lp d^{-1}\rp^{ab} r_{bK}\widetilde\om_a - \mathcal{F}_K\lp\widehat d^{-1}\rp^{\al\beta}\widehat r_{\beta K}\widetilde\m_\al.
\end{align}
Since there are no ${\mathbb Z}_4$ invariant 1-forms we have
immediately $\mathcal{W}_4=\mathcal{W}_5=0$. To determine
$\mathcal{W}_1$ we use the fact that $\mathcal{W}_2$ is primitive
and that $\int_X \omega_{a}\wedge \tilde{\omega}_{b} =
d_{ab}$.
\begin{align}
\int_X \text{d} \Omega \wedge J = -\mathcal{Z}_K r_{aK} v_a &= \int_X \mathcal{W}_1  J \wedge J \wedge J= \mathcal{W}_1 6 \mathcal{V}_6\\
\Rightarrow \, \mathcal{W}_1 &= -\frac{\mathcal{Z}_K r_{aK} v_a}{6
\mathcal{V}_6}.
\end{align}
Now we can read off $\mathcal{W}_3=\left( 2i
\mathcal{Z}_L r_{aL} v_a \mathcal{F}_K - r_{aK} v_a
\right) b_K$. It is straight forward to calculate $\mathcal{W}_2$. The torsion classes for a generic choice of metric fluxes are
\bea
\mathcal{W}_1 &=& -\frac{\mathcal{Z}_K r_{aK} v_a}{6\mathcal{V}_6},\\
\mathcal{W}_2 &=& -\lp\mathcal{W}_1v_a+\lp\k v\rp^{-1\,ab}\mathcal{Z}_Kr_{bK}\rp\om_a-\lp\widehat\k v\rp^{-1\,\al\beta}\mathcal{F}_K\widehat r_{\beta K}\m_\al,\\
\mathcal{W}_3 &=& \left( 2i \mathcal{Z}_L r_{aL}
v_a \mathcal{F}_K - r_{aK} v_a
\right) b_K,\\
\mathcal{W}_4 &=& \mathcal{W}_5 =0,
\eea
where $(\k v)^{-1}$ is the inverse of the matrix $(\k v)_{ab}=\k_{abc}v_c$, and similarly $(\widehat\k v)^{-1}$ is the inverse of the matrix $(\widehat\k v)_{\al\beta}=\widehat\k_{a\al\beta}v_a$.

Note that the twisted torus is generically not half-flat. For the twisted torus to be half-flat we would have to demand that $\Im(\mathcal{W}_1)=\Im(\mathcal{W}_2)=\mathcal{W}_4
=\mathcal{W}_5 =0$. This is equivalent to $\text{d} J \wedge J =
\text{d(Im}( \Omega)) =0$ or that $\Im({\mathcal F}_K)\widehat r_{\al K}=0$.  But these are precisely the D-term equations that we derived in section~\ref{D-terms}.  So solving the D-term equations is precisely equivalent to demanding that our manifold be half-flat.

Supersymmetric solutions should also have $\mathcal{W}_3=0$ \cite{Lust:2004ig}.  And indeed we can show this using the F-term equations (\ref{omegaImDKW}).  We have
\be
0=\mathcal{Z}_K\lp\Im D_KW\rp=2\mathcal{Z}_Kr_{aK}v_a+2e^D\Re W,\qquad\Longrightarrow\qquad\Re W=-e^{-D}\mathcal{Z}_Kr_{aK}v_a,
\ee
and then plugging this back in to (\ref{omegaImDKW}) we find that each component of $\mathcal{W}_3$ must vanish in a supersymmetric solution.

Note also that in \cite{Grana:2004bg} it was shown that Minkowski vacua of type IIA require $\mathcal{W}_1=0$.
This fits nicely with the observation that
$\mathcal{W}_1=e^D\Re W/(6\mathcal{V}_6)$.

So we see that our results agree very nicely with the language of $\SU(3)$ structure and torsion classes.

\section{Comparison of two different derivations of the Bianchi identities}
\label{BIDerivation}

Here we present two derivations of the Bianchi identities.

The usual derivation~\cite{Aldazabal:2006up} is to note that upon reducing on a $d$-dimensional torus we have (ignoring for now any orientifold group) $d$ vectors from reducing the metric and $d$ vectors from reducing the $B$-field.  Let $Z_i$ and $X^i$ respectively generate the gauge transformations for these two groups of vectors.  One then argues, by T-duality or otherwise, that the NS-NS fluxes must appear in the Lie brackets as
\bea
\ls Z_i,Z_j\rs &=& \om^k_{ij}Z_k-H_{ijk}X^k,\non\\
\ls Z_i,X^j\rs &=& -\om^j_{ik}X^k+Q^{jk}_iZ_k,\\
\ls X^i,X^j\rs &=& Q^{ij}_kX^k-R^{ijk}Z_k.\non
\eea
The Jacobi identities for this Lie algebra then give us the NS-NS Bianchi identities:
\bea
H_{k[i_1i_2}\om^k_{i_3i_4]} &=& 0,\non\\
H_{k[i_1i_2}Q^{k\,j}_{i_3]}+\om^j_{k[i_1}\om^k_{i_2i_3]} &=& 0,\non\\
H_{k\,i_1i_2}R^{k\,j_1j_2}+\om^k_{i_1i_2}Q^{j_1j_2}_k-4\om^{[j_1}_{k[i_1}Q^{j_2]k}_{i_2]} &=& 0,\label{AppendixBI}\\
\om^{[j_1}_{k\,i}R^{j_2j_3]k}+Q^{k[j_1}_iQ^{j_2j_3]}_k &=& 0,\non\\
Q^{[j_1j_2}_kR^{j_3j_4]k} &=& 0,\non
\eea

Let us present an alternative derivation, as suggested by~\cite{Shelton:2006fd}.  We have seen that it is natural to replace the exterior derivative $d$ acting on R-R forms by a covariant derivative $\mathcal{D}$.  We saw in section~\ref{EFTApproach} that such an object was what appeared in the tadpole condition and superpotential, and argued that it should also be used in finding the correct gauge transformations.  And in section~\ref{BFApproach} we saw that it could be understood as a covariant derivative for the spin bundle of which R-R fields formed sections.  By combining these considerations\footnote{From the base-fiber approach we did not require $\om^a_{ia}=0$, and can argue for the dependence on this trace, and from the effective field theory approach we can deduce how $R$ must appear.}, it is natural to define the general $\mathcal{D}$ by its action on a $p$-form as
\begin{align}
\mathcal{D}A^{(p)} =& \binom{p+3}{3}H_{[i_1i_2i_3}A_{i_4\cdots i_{p+3}]}\,\frac{1}{(p+3)!}dx^{i_1}\w\cdots\w dx^{i_{p+3}}\non\\
& -\left\{\binom{p+1}{2}\om^j_{[i_1i_2}A_{|j|i_3\cdots i_{p+1}]}+\frac{p+1}{2}\om^j_{j[i_1}A_{i_2\cdots i_{p+1}]}\right\}\frac{1}{(p+1)!}dx^{i_1}\w\cdots\w dx^{i_{p+1}}\non\\
& +\hlf\left\{\binom{p-1}{1}Q^{jk}_{[i_1}A_{|jk|i_2\cdots i_{p-1}]}+\binom{p-1}{0}Q^{jk}_jA_{k\,i_1\cdots i_{p-1}}\right\}\frac{1}{(p-1)!}dx^{i_1}\w\cdots\w dx^{i_{p-1}}\non\\
& -\frac{1}{6}\binom{p-3}{0}R^{jk\ell}A_{jk\ell\,i_1\cdots i_{p-3}}\,\frac{1}{(p-3)!}dx^{i_1}\w\cdots\w dx^{i_{p-3}}.
\label{GeneralD}
\end{align}

For consistency, we need $\mathcal{D}$ to share a key property with the exterior derivative that it is replacing, namely that $\mathcal{D}^2=0$ on all forms.  Computing,
\begin{align}
\mathcal{D}^2A^{(p)} &= -6\binom{p+4}{4}H_{k\,i_1i_2}\om^k_{i_3i_4}A_{i_5\cdots i_{p+4}}\,\frac{1}{(p+4)!}dx^{i_1}\w\cdots\w dx^{i_{p+4}}\non\\
& +\left\{-\binom{p+2}{2}\lp H_{k\ell\,i_1}Q^{k\ell}_{i_2}-\hlf Q^{k\ell}_kH_{\ell\,i_1i_2}-\hlf \om^k_{k\ell}\om^\ell_{i_1i_2}\rp A_{i_3\cdots i_{p+2}}\right.\non\\
& \quad\left.+3\binom{p+2}{3}\lp H_{k\,i_1i_2}Q^{kj}_{i_3}+\om^k_{i_1i_2}\om^j_{k\,i_3}\rp A_{j\,i_4\cdots i_{p+2}}\right\}\frac{1}{(p+2)!}dx^{i_1}\w\cdots\w dx^{i_{p+2}}\non\\
& +\left\{\binom{p}{0}\lp-\frac{1}{6}H_{k\ell m}R^{k\ell m}+\frac{1}{4}\om^k_{k\ell}Q^{\ell m}_m\rp A_{i_1\cdots i_p}\right.\non\\
& \quad\left.+\hlf\binom{p}{1}\lp H_{k\ell\,i_1}R^{k\ell j}-Q^{k\ell}_{i_1}\om^j_{k\ell}-\om^k_{k\ell}Q^{\ell j}_{i_1}-Q^{k\ell}_k\om^j_{\ell\,i_1}\rp A_{j\,i_2\cdots i_p}\right.\\
& \quad\left.-\hlf\binom{p}{2}\lp H_{k\,i_1i_2}R^{k\,j_1j_2}+4\om^{j_1}_{k\,i_1}Q^{k\,j_2}_{i_2}+Q^{j_1j_2}_k\om^k_{i_1i_2}\rp A_{j_1j_2\,i_3\cdots i_p}\right\}\frac{1}{p!}dx^{i_1}\w\cdots\w dx^{i_p}\non\\
& +\left\{\hlf\binom{p-2}{0}\lp\om^{j_1}_{k\ell}R^{k\ell\,j_2}+\hlf\om^k_{k\ell}R^{\ell\,j_1j_2}+\hlf Q^{k\ell}_kQ^{j_1j_2}_\ell\rp A_{j_1j_2\,i_1\cdots i_{p-2}}\right.\non\\
& \quad\left.+\hlf\binom{p-2}{1}\lp R^{k\,j_1j_2}\om^{j_3}_{k\,i_1}+Q^{j_1j_2}_kQ^{k\,j_3}_{i_1}\rp A_{j_1j_2j_3\,i_2\cdots i_{p-2}}\right\}\frac{1}{(p-2)!}dx^{i_1}\w\cdots\w dx^{i_{p-2}}\non\\
& -\frac{1}{4}\binom{p-4}{0}Q^{j_1j_2}_kR^{k\,j_3j_4}A_{j_1j_2j_3j_4\,i_1\cdots i_{p-4}}\,\frac{1}{(p-4)!}dx^{i_1}\w\cdots\w dx^{i_{p-4}}.\non
\end{align}

From this we find that in order to ensure that $\mathcal{D}^2=0$ on all forms, we need precisely the Bianchi identities found above, and one additional one which does not follow by contraction,
\be
2H_{k\ell m}R^{k\ell m}+3\om^k_{k\ell}Q^{m\ell}_m=0.
\ee
Note that this final identity is satisfied on {\it any} IIA orientifold, because there are generally no scalars (zero-forms) which are odd under the involution.

The coefficients of the two trace terms in (\ref{GeneralD}) can be argued from the spin bundle transformations of R-R fields, as in section~\ref{GeneralSituation}, along with a T-duality argument to get the $\Tr Q$ term in terms of the $\Tr\om$ term, but there is another nice check as well.  If the coefficients of the trace terms were at all different, then $\mathcal{D}^2=0$ would lead to more constraints beyond the single extra constraint we found above.  Though not inconsistent, these additional requirements seem surprising and ad-hoc.  With the given coefficients, however, these additional constraints follow simply from the traces of constraints with more free indices.  

%It would be gratifying to better understand the relation between the two derivations presented here.

%\bibliography{branes}
\providecommand{\href}[2]{#2}\begingroup\raggedright\endgroup

\bibliographystyle{utphys}
\end{document}